\documentclass[preprint,aps,superscriptaddress,floatfix,preprintnumbers,
altaffilletter]{revtex4-1}
\usepackage[utf8]{inputenc}
\usepackage[colorlinks=true,citecolor=blue,linkcolor=blue]{hyperref}
\usepackage{amsmath,amssymb}
\usepackage{epsfig}
\usepackage{graphicx}             
\usepackage{url}
\usepackage{color}
\usepackage{slashed}
\usepackage{multirow}
\usepackage{placeins}
\usepackage[dvipsnames]{xcolor}
\usepackage{epstopdf}
\usepackage[utf8]{inputenc}
\usepackage[font=small,labelfont=bf]{caption}
\usepackage{tikz}
\usetikzlibrary{trees}
\usetikzlibrary{decorations.pathmorphing}
\usetikzlibrary{decorations.markings}

\usepackage{cleveref}
\usepackage{subfigure} 
\usepackage{caption}
\usepackage{lipsum}

\newcommand{\beq}{\begin{equation}}
\newcommand{\eeq}{\end{equation}}

\newcommand{\GeV}{\, \text{GeV}}

\def\mA{\mathcal{A}}

\def\mL{\mathcal{L}}

\def\mO{\mathcal{O}}

\begin{document}

\title{Bottom-quark Forward-Backward Asymmetry,\\ Dark Matter and the LHC}


\author{Da Liu}
\affiliation{\mbox{High Energy Physics Division, Argonne National Laboratory, Argonne, IL 60439}}

\author{Jia Liu}
\affiliation{\mbox{Physics Department and Enrico Fermi Institute, University of Chicago, Chicago, IL 60637}}

\author{Carlos E.M. Wagner}
\affiliation{\mbox{High Energy Physics Division, Argonne National Laboratory, Argonne, IL 60439}}
\affiliation{\mbox{Physics Department and Enrico Fermi Institute, University of Chicago, Chicago, IL 60637}}
\affiliation{\mbox{Kavli Institute for Cosmological Physics, University of Chicago, Chicago, IL 60637}}

\author{Xiao-Ping Wang}
\affiliation{\mbox{High Energy Physics Division, Argonne National Laboratory, Argonne, IL 60439}}

\date{\today}
\preprint{EFI-17-26}

\begin{abstract}
The LEP experiment at CERN provided accurate measurements of the $Z$ neutral gauge boson properties. Although all 
measurements agree well with the SM predictions, the forward backward asymmetry of the bottom-quark remains almost
3$\sigma$ away from the SM value. We proposed that this anomaly may be explained by the existence of a new $U(1)_D$
gauge boson, which couples with opposite charges to the right-handed components of the bottom and charm quarks.   Cancellation
of gauge anomalies demands the presence of a vector-like singlet charged lepton as well as a neutral Dirac (or Majorana) particle
that provides a Dark Matter candidate. Constraints from precision measurements imply that the mass of the new gauge boson
should be around $115$~GeV.  We discuss the experimental constraints on this scenario, including the existence of a di-jet resonance 
excess at an invariant mass similar to the mass of this new gauge boson, observed in boosted topologies at the CMS experiment.

\end{abstract}

\maketitle

\setcounter{secnumdepth}{2} 
\setcounter{tocdepth}{2}   

\tableofcontents

\section{Introduction}
The Standard Model (SM) provides an accurate description of all experimental observables. The discovery of a 125~GeV resonance
with properties consistent with a 125~GeV Higgs boson~\cite{Aad:2012tfa,Chatrchyan:2012xdj}
provides evidence of the realization of the Higgs mechanism as a source of gauge boson and fermion masses.  However, the exact
properties of the Higgs sector are still unknown.  The minimal model postulates the existence of just one Higgs,
transforming as a doublet under the gauge interactions. Precision measurements of the charged and neutral gauge boson 
properties~\cite{ALEPH:2005ab}
show the preference towards a doublet Higgs state.  Similar properties would be obtained, however, if there were more than just
one Higgs doublets. Finally, the presence of extra singlet scalar Higgs states is not constrained by these considerations. 

Another outstanding question is the origin of the Dark Matter (DM) observed in astrophysical configurations. The Standard Model does
not provide any DM candidate and its nature is unknown. Among the many DM candidates, weakly interactive massive particles (WIMPs)
are particularly attractive since they can easily be incorporated in beyond the SM scenarios. Moreover, it is well known that WIMPs
with mass of the order of the weak scale and interactions of about the weak scale one provide a good candidate of thermal DM 
candidate~\cite{Bertone:2004pz}.

Precision measurements of the gauge sector have shown agreement with expected SM properties at the per-mille level. Such 
a precision leads to sensitivity to radiative corrections which depend in a relevant way on the top-quark and the Higgs mass. 
Among the many observables measured, the bottom forward-backward asymmetry measured at LEP presents a 3~$\sigma$ deviation
with respect to the values expected in the SM~\cite{ALEPH:2005ab}.  Although this deviation could be just due to statistical fluctuations, its nature is intriguing
since it could be associated with a large correction to the right-handed bottom quark coupling to the $Z$ boson, which may only
be explained by either mixing of the bottom-quark with additional (vector-like) quarks, or by mixing of the $Z$ gauge boson with
additional neutral gauge bosons.  The first possibility  led to the proposal of what are  called Beautiful-Mirror scenarios~\cite{Choudhury:2001hs}, and their properties have been studied in detail~\cite{Morrissey:2003sc, Kumar:2010vx, Batell:2012ca}.  
The second possibility, namely the existence of additional gauge bosons contributing via
mixing to a variation of the bottom quark coupling has also been explored, within the context of left-right models and warped 
extra dimensions~\cite{He:2002ha,Djouadi:2006rk}. In this article, we study the properties of a neutral gauge boson with preferential
couplings to the bottom and charm quarks. We
shall show that it leads naturally to the existence of a low energy spectrum that includes two Higgs doublets, a singlet, and
a charged and a neutral vector-like singlets, the latter being a good DM candidate. 

This article is organized as follows. In section~\ref{sec:Model}, we describe the properties of the proposed SM gauge extension. 
We present the tree-level couplings of the new gauge boson to SM particles, as well as the necessary fermion content in order
to cancel the gauge anomalies. The new Higgs bosons are introduced in order to induce the necessary mixing and provide
masses to all chiral fermions in the theory.  In section~\ref{sec:EWPT}, we study the constraints on this model coming from
precision electroweak measurements. In section~\ref{sec:K-Production}, we study the collider constraints on this model and
in section~\ref{sec:DM-search} we study the constraints coming from
the requirement of obtaining the proper DM relic density without being in conflict with direct and indirect
detection constraints. We reserve section~\ref{sec:Conclusions} to our conclusions.

\section{ A Model with Two Higgs Doublets and A Singlet}
\label{sec:Model}

In this section, we shall describe the precise gauge extension of the SM we propose to explain the anomalous value of the bottom-quark forward-backward asymmetry.  We consider  a new gauge group $U(1)_D$ with gauge boson field $K_\mu$~\cite{Liu:2017lpo}, under which, the right-handed bottom and right-handed charm quark have opposite charge $\pm X$. This ensures  the automatic cancellation of the  $SU(3)_c^2\times U(1)_D, U(1)_D^3$  gauge anomaly. In order to cancel the gauge anomalies involving the hyper-charge gauge field, we introduce two $SU(2)$ singlet SM-vector-like  leptons $\chi_{1,2}$ with hyper charge -1 and 0, where only the right-handed components are charged under $U(1)_D$, carrying charges $\pm X$, respectively.   The neutral state $\chi_2$ will be  naturally a dark matter candidate,  provided we impose a $Z_2$ parity, under which $\chi_2$ transforms non-trivially while SM-particles are neutral under this symmetry transformations. 

A modification of the forward-backward asymmetry, consistent with the one observed experimentally, may be obtained by a sizable variation of the coupling of the $Z$ to right-handed bottom quarks~\cite{Choudhury:2001hs}. Such a  variation 
of the $Z$ gauge boson couplings may be the result of mixing between the $Z$ and the $K$ gauge bosons. Such mixing may be induced 
by a new $SU(2)$  Higgs doublet $\Phi_1$ with hyper-charge $Y = 1/2$ and $U(1)_D$ charge equal to the $b_R$ one, which  is needed to make sure that we obtain the enhanced $Zb_R\bar{b}_R$ coupling for $m_K > m_Z$.  

The SM Higgs-like doublet which gives the other SM fermions and the gauge bosons  masses will be denoted as $\Phi_2$. Another SM gauge singlet scalar $\Phi_3$ charged under $U(1)_D$ is needed to give  mass to the $K$ gauge boson. It is clear that within this setup, we can not write down the normal Yukawa interaction for the bottom and charm quark directly. To solve the problem, we add  two vector-like quarks  $\psi_b$, $\psi_c$, which have the same SM charges as $b_R$ and $c_R$, but without $U(1)_D$ charge. The masses of the bottom and charm quarks  are obtained by their mixing with the heavy vector-like quarks, which is in the same spirit of partial compositeness~\cite{Kaplan:1991dc}. The particle contents of our model and their gauge group charges are listed in Table~\ref{tab:gaugecharge}.

\begin{table}[h]
\begin{tabular}{|c||c|c|c||c||}
\hline 
filed & $SU(3)_C$ &$SU(2)_L$  &$U(1)_Y$ & $U(1)_D$ \\  \hline
\hline
 $\Phi_1$ & 1 & 2   &  $\frac{1}{2}$ & X \\
 \hline
 $\Phi_2$  & 1   & 2  & $\frac{1}{2}$  & 0   \\
 \hline
 $\Phi_3$  &  1   & 1  &  0                  &  -X    \\
\hline
$b_R$   &  3  & 1   & $-\frac{1}{3}$    & X   \\
\hline
$c_R$   &   3   & 1  & $\frac{2}{3}$ & -X   \\  \hline
$\chi_{1,R}$  & 1  & 1  &  -1   & X  \\  \hline
$\chi_{2,R}$  & 1  & 1  &  0   &  -X  \\  \hline
$\chi_{1,L}$  & 1  & 1  &  -1   & 0    \\  \hline
$\chi_{2,L}$  & 1  & 1  &  0   &  0  \\  \hline
$\psi _{b}$  & 3  & 1  &  $-\frac{1}{3}$  & 0  \\
\hline
$\psi _{c}$  & 3  & 1  &  $\frac{2}{3}$  & 0  \\
\hline
\end{tabular}
\caption{All particles with $SU(3)_C\times SU(2)_L \times U(1)_Y \times U(1)_D$ charge
specified. The anomaly-free condition is applied for this model. The $U(1)_D$ charge of
$b_R$, $c_R$, $\chi_{1,R}$ and $\chi_{2,R}$ is determined by the anomaly-free condition. 
We choose $X = 1$ for the model without loss of generality.
}
 \label{tab:gaugecharge}
\end{table}

The whole Lagrangian in our models can be written into three parts:
\begin{align}
\mathcal{L}=  \mL_\Phi + \mL_q + \mL_\ell, 
\end{align}
where $\mL_{\Phi, q, \ell}$ denotes that Lagrangian in the Higgs sector, the quark sector and the lepton sector respectively. For the Higgs part, the Lagrangian is simply as follows:
\begin{align}
\mathcal{L}_\Phi=   |D_\mu \Phi_1|^2 + |D_\mu \Phi_2 |^2 +|D_\mu \Phi_3|^3 - V(\Phi_i, \Phi_i^\dagger) ,
\end{align}
where the covariant derivative is defined as:
\beq
D_\mu = \partial_\mu - i g_D Y_D K_\mu - i g W_\mu^a \frac{\sigma^a}{2} - i g' Y B_\mu ,
\eeq 
where $K_\mu$ is the $U(1)_X$ gauge boson, $W_\mu^a$ are the SM $SU(2)_L$ gauge bosons and $B_\mu$ is the $U(1)_Y$ 
hypercharge gauge boson. The gauge bosons denoted without tildes are gauge eigenstates. After considering mixing effects, we
shall later use tildes to denote mass eigenstates. For $W^\pm_\mu$ and the photon $A_\mu$, since they do not mix with $K_{\mu}$, the notation is the same as in the SM  and there is no need to add tildes.  
The Higgs potential will be fully discussed in next subsection, and here we just assume that the fields associated with the 
three CP-even neutral Higgs bosons obtain vacuum expectation values (vev), i.e.
\begin{align}
\begin{array}{ccc}
\Phi_1=\left(\begin{array}{c}
h_1^+\\
\frac{1}{\sqrt{2}}\left(v_1+h_1^0+ia_1^0\right)
\end{array}\right), &  $~~~$
\Phi_2=\left(\begin{array}{c}
h_2^+\\
\frac{1}{\sqrt{2}}\left(v_2+h_2^0+ia_2^0\right)
\end{array}\right), &  $~~~$
\Phi_3=\frac{1}{\sqrt{2}}\left(v_D+ h_3^0 + ia_3^0\right) \,.  \\
\end{array}
\label{eq-HVeV}
\end{align}
The vev's do not break the electromagnetism symmetry, and $\Phi_1$ induce the mixing between the neutral massive gauge bosons $K_\mu$ and $Z_\mu$, which are proportional to $v_1^2$. Since the $W$ boson mass is not modified, the custodial symmetry is explicitly broken by the mixing and this will be reflected in $T$ parameter. The high-precision constraints on the $T$ parameter  tell us that the mixing should be very small, which favors a small vev, $v_1 \ll v_{2,D}$. For later convenience, it is useful to define the ratio 
angle $\beta$:
\beq
\tan\beta = \frac{v_2}{v_1},
\eeq
which controls the charged Higgs mixing by Goldstone equivalence theorem and has to be large. In this limit, the neutral CP-even Higgs $h_2^0$ will roughly be the SM-like 125 GeV Higgs boson observed at the LHC~\cite{Aad:2012tfa,Khachatryan:2016vau}, and mixes with the CP-even Higgs boson $h_1^0$. The remaining physical charged Higgs and CP-odd Higgs bosons will be $\Phi_1$-like, while $h_2^+, a_2^0, a_3^0$ becomes the dominant longitudinal part of the massive $W, Z, K$ gauge bosons. $\Phi_1$-like physical Higgs will couple to SM gauge bosons and fermions suppressed by mixing angle to SM-like Higgs $\cot\beta$ . The last CP-even Higgs boson will be $\Phi_3$-like and only couple largely to $U(1)_D$ charged particles and the $U(1)_D$ gauge boson  $K_\mu$. As its vev $v_D$ is the source of bottom and charm masses, it couples with them proportional to their masses, i.e. ${m_{b,c}/v_D}$.  In the absence of mixing with the other CP-even states it will be produced in bottom-fusion and gluon fusion processes and it will decay mostly to bottom quarks. Hence, provided the mixing with the SM-like Higgs boson is small, the LHC constraints on it are expected to be very weak.

The most general interactions in the quark sector are given by:
\begin{align}
\mathcal{L}_q & = \sum_{q } i \bar{q} \slashed D q - m_{b,\psi}  \bar{\psi}_{b} \psi_{b} - m_{c,\psi}  \bar{\psi}_{c} \psi_{c} - \left( \bar{Q}^i_L y^{ij}_{2u} \tilde{\Phi}_2u_R^j
+ \bar{Q}_L^i y_{2d}^{ij} \Phi_2 d_R^{j} + h.c.\right)  \nonumber\\ 
&-y^i_{2b,\psi} \bar{Q}_L^{i} \Phi_2 \psi_{b,R} -  y^i_{2c,\psi}   \bar{Q}_L^{i}\tilde \Phi_2 \psi_{c,R}  - y_{3b,\psi}\bar{\psi}_{b,L}  \Phi_3 b_{R}  - y_{3c,\psi}\bar{\psi}_{c,L}  \Phi_3^* c_{R}  +h.c \,,
\end{align}
where  $\tilde{\Phi}_2=i\sigma_2 \Phi_2^\dag$, $Q_L^i$ is the three family SM $SU(2)$ quark doublet and $u_R^j = (u_R, t_R), d_R^j = (d_R, s_R)$.  The vev of $\Phi_3$ will induce the mixing between the right-handed bottom and charm quarks, $b_R,c_R$, and their  corresponding vector-like quark partner. As a result, the bottom and charm quarks  obtain masses after Electroweak spontaneous Symmetry Breaking (EWSB). In this sense, it is very similar to the partial compositeness scenario of the composite Higgs models except that our vector-like quark partners can be fundamental. It is not difficult to embed our model to a composite Higgs model, where all the Higgs bosons are Goldstone bosons associated with spontaneously broken global symmetry of a new strong sector.  

As described above,  the masses of the bottom and charm quark arise from the spontaneously broken $U(1)_D$ gauge symmetry and electroweak gauge symmetry, which can also been seen by integrating out the heavy vector-like quark $\psi_{b,c}$ at the tree level using equation of motion:
\begin{eqnarray}
\psi_{b,R} & = & - \frac{y_{3b,\psi}}{m_{b,\psi}} \Phi_3 b_R + \cdots, \qquad  \psi_{b,L} = - \frac{y^i_{2b,\psi}}{m_{b,\psi}} \Phi_2^\dagger Q_L^i + \cdots, 
\nonumber\\
\psi_{c,R} & = & - \frac{y_{3c,\psi}}{m_{c,\psi}} \Phi_3^* b_R + \cdots, \qquad  \psi_{c,L} = - \frac{y^i_{2c,\psi}}{m_{c,\psi}} \Phi_2^\dagger Q_L^i + \cdots ,
\end{eqnarray}
then we have the effective Yukawa interaction Lagrangian:
\begin{align}
\mathcal{L}_q^Y & =  
- \left( \bar{Q}^i_L y^{ij}_{2u} \tilde{\Phi}_2u_R^j
+ \bar{Q}_L^i y_{2d}^{ij} \Phi_2 d_R^{j}  + \frac{ y^i_{2b,\psi} y_{3b,\psi}}{m_{b,\psi}}\bar{Q}_L^{i} \Phi_2    \Phi_3 b_R+ \frac{ y^i_{2c,\psi} y_{3c,\psi}}{m_{c,\psi}}\  \bar{Q}_L^{i}\tilde \Phi_2 \Phi_3^* c_{R}   +h.c \right) \,,
\label{eq:YukEff}
\end{align}
It is clear that the flavor  interaction structure of $h_2^0$ is of SM-like and the effective Yukawa couplings may be diagonalized at the same time as the mass matrices. Although the last two terms in Eq.~(\ref{eq:YukEff}) can in principle induce flavor changing neutral current (FCNC) in the quark sector, it is very model dependent. In the following, we will assume the flavor-off-diagonal  interactions are very small, which is equivalent to start with the Lagrangian with following parameters:
\beq
y_{2b,\psi}^i = \delta^{i3}y_{2b,\psi}, \qquad y_{2c,\psi}^i = \delta^{i2}y_{2c,\psi}, \qquad y^{2j}_{2u} =0 ,\qquad  
 y^{3j}_{2d} =0.
\eeq

For the leptons, we will focus on the third generation and similarly neglect off-diagonal terms between different generations. The Yukawa interaction Lagrangian reads:
 \begin{align}
\mathcal{L}_\ell^Y = 
- y_{\tau}\bar{L}_{L,\tau}\Phi_2\tau_{R}-y_{\chi_1}\bar{\chi}_{1,L}\Phi_3\chi_{1,R} -y_{\chi_2}\bar{\chi}_{2,L}\Phi_3^*\chi_{2,R} - \frac12 M_m \bar{\chi}_{2,L} \chi_{2,L}^c - m_{\tau_1} \bar{\chi}_{1,L} \tau_R + h.c.,
\label{eq:YukawaChis}
\end{align}
where we have imposed the $Z_2$ parity for the neutral lepton $\chi_2 \rightarrow - \chi_2$ and assumed that  $\chi_1$ only mix with the third generation charged lepton $\tau_R$ by  the direct Dirac mass $m_{\tau_1}$, which is the only source of  $\chi_1$ decay.

\subsection{The  gauge  sector}
In this subsection, we will discuss the mixing in the gauge sector and the couplings of the dark gauge boson. 
After the gauge symmetry breaking, the charged gauge boson sector is the same as SM with $v^2 = v_1^2 + v_2^2$:
\beq
m_W^2 = \frac14 g^2 v^2.
\eeq
For the neutral sector, we first apply the rotation to transform ${W_3, B}$ gauge bosons into 	${Z, A}$ gauge bosons as in the SM.
The $\Phi_1$ is charged under both SM $SU(2)_L \times U(1)_Y$ and  $U(1)_D$, thus
induces off-diagonal mass terms for $Z_\mu$ and $K_\mu$, but the photon state $A_\mu$ is not affected and stays massless, as it should be. 
Factoring out the photon state $A_\mu$, the $Z_\mu$ and $K_\mu$ will mix with each other and the mixing mass-square matrix is 
given by
\begin{align}
M_{V}^2=\left(\begin{array}{cc}
m_Z^2 & - \frac{2 g_D c_\beta^2 }{\sqrt{g^2 + g'^2}} m_Z^2 \\
- \frac{2 g_D c_\beta^2 }{\sqrt{g^2 + g'^2}}  m_Z^2 & 
m_K^2 +  \frac{4 g_D^2 c_\beta^2}{g^2 + g'^2}  m_Z^2\\
\end{array}\right),
\end{align}
where we have defined:
\beq
m_Z^2=\frac{(g^2+g'^2)v^2}{4}, \qquad c_\beta = \cos\beta= \frac{v_1}{v},  \qquad m_K^2=g_D^2 v_D^2.
\eeq
The matrix can be easily diagonalized by an $2\times 2$ orthogonal matrix with mixing angle $\alpha$:
\begin{align}
\left(\begin{array}{c}
Z_\mu\\
K_\mu
\end{array}\right) = 
\left(\begin{array}{cc}
\cos\alpha &  \sin\alpha\\
-\sin\alpha &\cos\alpha
\end{array}\right)
\left(\begin{array}{c} \tilde{Z}_\mu \\ \tilde{K}_\mu   \end{array}\right) ,
\end{align}
where ${\tilde{Z}_\mu, \tilde{K}}_\mu$ are the final mass eigenstates. 
As will be discussed in detail in Sec.~\ref{sec:EWPT}, the Electroweak precision test (EWPT), including the T parameter and Z-pole measurements, put a strong constraint on the mixing angle thus the mixing should be very small,  which further indicates  $c_\beta^2 \ll 1$.  Then the value of $\sin\alpha$ can be approximately given by:
\begin{align}
\sin\alpha  \sim- \frac{2 g_D c_\beta^2}{\sqrt{g^2 + g'^2}}  \frac{m_Z^2}{m_K^2 - m_Z^2},
\label{eq:sinalpha}
\end{align}
where we have kept the leading terms in a $c_\beta^2$ expansion.

The mass eigenvalues of the gauge bosons are simply:
\begin{align}
m^2_{\tilde{Z}} & \approx m_Z^2 - \sin^2\alpha (m_K^2 - m_Z^2)+ \mathcal{O}(\sin^3\alpha)\\
m^2_{\tilde{K}} & \approx m_K^2 + \frac{4 g_D^2 c_\beta^2}{g^2+g'^2}  m_Z^2 
+ \sin^2\alpha (m_K^2 - m_Z^2)
+\mathcal{O}(\sin^3\alpha),
\end{align}
Due to the mixing between $\tilde K$ and $\tilde Z$, the coupling of $\tilde Z$ to SM particles and also the $\tilde Z$ mass will be modified with respect to their SM values. We will carefully discuss it afterwards. At 1-loop level, the kinetic mixing term $\epsilon K_{\mu\nu} B^{\mu\nu}$ 
can be induced from the fermions which charged under both $U(1)_Y$ and $U(1)_D$, with $\epsilon \sim g_D g'/(16 \pi^2)$.
Given it is much smaller than the direct mixing $\sin \alpha$ from vev of $\Phi_1$, we can neglect this term.

\subsection{Higgs sector}
In this subsection, we will discuss the Higgs sector and get the mass eigenstates of Higgs. First, we write down the 
general scalar potential which is gauge invariant under 
$SU(3)_C\times SU(2)_L \times U(1)_Y\times U(1)_D$ as follows:
\begin{align}
V& =\mu_1^2\Phi_1^\dag\Phi_1 +\mu_2^2\Phi_2^\dag \Phi_2 +\mu_3^2\Phi_3^\dag \Phi_3 \\ \nonumber
&+\lambda_1(\Phi_1^\dag\Phi_1)^2+\lambda_2(\Phi_2^\dag\Phi_2)^2 +\lambda_3(\Phi_3^\dag\Phi_3)^2\\ \nonumber &+\lambda_4(\Phi_1^\dag\Phi_1)(\Phi_2^\dag \Phi_2) +\lambda_5(\Phi_1^\dag\Phi_1)(\Phi_3^\dag\Phi_3) +\lambda_6(\Phi_2^\dag\Phi_2)(\Phi_3^\dag\Phi_3) +\lambda_7(\Phi_1^\dag\Phi_2)(\Phi_2^\dag\Phi_1)\\ \nonumber
&+\mu_8\left(\Phi_1^\dag \Phi_2 \Phi_3^{*} + h.c.\right) \,,
\end{align}
The minimum condition of $V$ can be always satisfied by requiring the mass terms have the following relationship
\begin{align}
\mu_1^2&=-\left(\lambda_1 v_1^2+\frac{\lambda_4+\lambda_7}{2}v_2^2+\frac{\lambda_5}{2} v_D^2+\mu_8\frac{v_2 v_D}{\sqrt{2}v_1}\right), \\ \nonumber
\mu_2^2&=-\left(\lambda_2 v_2^2+\frac{\lambda_4+\lambda_7}{2}v_1^2+\frac{\lambda_6}{2} v_D^2+\mu_8\frac{v_1 v_D}{\sqrt{2}v_2}\right), \\ \nonumber
\mu_3^2&=-\left(\lambda_3 v_D^2 +\frac{1}{2}(\lambda_5 v_1^2+\lambda_6 v_2^2)+\mu_8\frac{v_1v_2}{\sqrt{2}v_D},\right),
\end{align}
where the vevs of the Higgs are defined in Eq.~(\ref{eq-HVeV}). Let's start from the charged Higgs mass matrix, which is straightforward to obtain by the second derivative of the potential $V$:
\begin{align}
M_{\pm}^2
=-\left(\frac{\lambda_7}{2}+\mu_8\frac{v_D}{\sqrt{2}v_1v_2}\right)
\left(\begin{array}{cc}
v_2^2 & -v_1 v_2\\
-v_1 v_2 & v_1^2
\end{array}\right).
\end{align}
The mass of the physical charged Higgs is:
\begin{align}
m^2_{H^\pm}= -\frac{\mu_8   v_D}{\sqrt{2}\sin\beta\cos\beta}  -\frac{\lambda_7v^2}{2}.
\end{align}
The two charged Higgs fields $h^\pm_1$ and $h^\pm_2$ mix to form the mass eigenstates $H^\pm$ and $G^\pm$ according to
\begin{align}
\left(\begin{array}{c} h^\pm_1 \\ h^\pm_2 \end{array}\right)=\left(\begin{array}{cc}
\sin\beta & \cos\beta \\
-\cos\beta   & \sin\beta \end{array}
\right)\left(\begin{array}{c} H^\pm \\ G^\pm \end{array}\right).
\end{align}

Similarly, we can obtain the mass eigenvalue of physical  CP-odd  Higgs as the trace of the mass matrix:
\begin{align}
M^2_{{\rm odd}}=-\frac{\mu_8}{\sqrt{2}}\left(
\begin{array}{ccc}
\frac{v_2v_D}{v_1}  &-v_D & v_2\\
-v_D  &  \frac{v_1v_D}{v_2}  & -v_1\\
v_2& -v_1 & \frac{v_1v_2}{v_D}
\end{array}\right),
\label{eq:massMCPodd}
\end{align}
whose value is given by:
\begin{align}
m^2_{A_0}= -\frac{\mu_8   v_D}{\sqrt{2}s_\beta  c_\beta}  -\frac{\mu_8 v^2 s_\beta c_\beta}{\sqrt{2}v_D},
\label{eq:massofCPodd}
\end{align}
where we have abbreviated $s_\beta \equiv \sin\beta, c_\beta \equiv \cos\beta$. The mass mixing matrix is listed in Appendix \ref{sec:CPoddHiggs}. From the masses, we can easily see that in the large $t_\beta$ limit, which is required by the small $\tilde K, \tilde Z$ mixing, the mass scales of the heavy charged Higgs and CP-odd Higgs can be as  large as TeV if $\mu_8$ is around the electroweak scale. In this limit, both heavy charged Higgs and CP-odd Higgs dominantly come
from $\Phi_1$.

Finally we consider the CP-even sector, which involves three physical states. The mass matrix is obtained as follows:
\begin{align}
M^2_{{\rm even}}=\left(\begin{array}{ccc}
2\lambda_1 v_1^2-\frac{\mu_8 v_2v_D}{\sqrt{2}v_1} & v_1v_2(\lambda_4+\lambda_7)+\frac{\mu_8 v_D}{\sqrt{2}} & \lambda_5 v_1v_D + \frac{\mu_8 v_2}{\sqrt{2}}\\
v_1v_2(\lambda_4+\lambda_7)+\frac{\mu_8 v_D}{\sqrt{2}} &2 \lambda_2v_2^2-\frac{\mu_8 v_1v_D}{\sqrt{2}v_2} & \lambda_6 v_2v_D+\frac{\mu_8 v_1}{\sqrt{2}}\\
 \lambda_5 v_1v_D + \frac{\mu_8 v_2}{\sqrt{2}}&\lambda_6 v_2v_D+\frac{\mu_8 v_1}{\sqrt{2}} & 2\lambda_3v_D^2 -\frac{\mu_8 v_1v_2}{\sqrt{2}v_D}
\end{array}\right).
\end{align}
As discussed before, in order to decouple the heavy charged Higgs and not induce the large mixing between SM Higgs and the other CP-even Higgs , we require that $\mu_8$ is roughly of $\mO(v_2)$ and $c_\beta \ll 1$. In order not to induce large mixing between the SM Higgs $h_2$ and the singlet $h_3$, we further require that $\lambda_6$ is small and of the same order as $c_\beta$. Under the above assumption, we can  simplify the mass matrix by eliminating
the quadratic and linear term of $v_1$, except $v_1 \mu_8$ terms, which since $v_D$ is of the same order as $v$,
are of the same order as $\lambda_6 v_2 v_D$. This is
equivalent to set $\lambda_1, \lambda_4, \lambda_5, \lambda_7$ to 0 and the CP even mass matrix is now: 
\begin{align}
\label{eq:higgsmm}
M^2_{{\rm even}}=\left(\begin{array}{ccc}
-\frac{\mu_8 v_2v_D}{\sqrt{2}v_1} & \frac{\mu_8 v_D}{\sqrt{2}} &  \frac{\mu_8 v_2}{\sqrt{2}}\\
* &2 \lambda_2v_2^2-\frac{\mu_8 v_1v_D}{\sqrt{2}v_2} & \lambda_6 v_2v_D+\frac{\mu_8 v_1}{\sqrt{2}}\\
 * & * & 2\lambda_3v_D^2 -\frac{\mu_8 v_1v_2}{\sqrt{2}v_D}
\end{array}\right).
\end{align}
The mass  eigenvalues  at leading order in $\cot \beta$ and $\lambda_6$ are simply as following:
\begin{align}
m^2_{H^0_1} &\simeq -\frac{\mu_8 v_D\tan\beta}{\sqrt{2} } \simeq m^2_{A^0}  ,\nonumber \\
m^2_{H^0_2} & \simeq 2 \lambda_2 v_2^2 ,
   \nonumber  \\
m^2_{H^0_3} & \simeq 2 \lambda_3 v_D^2 .\,
\label{eq:M2simply}
\end{align}
The unitary mixing  matrix is define as:
\begin{align}
\left(\begin{array}{c} h_1^0 \\ h^0_2 \\ h^0_3 \end{array}\right)   =
\left(\begin{array}{ccc}
U_{11} & U_{12}& U_{13}
\\
U_{21} & U_{22}& U_{23}\\
U_{31} & U_{32}& U_{33}\end{array}\right) 
\left(\begin{array}{c} H_1^0 \\ H^0_2 \\ H^0_3  
\end{array}\right),
\end{align}
where $h$ ($H$) denote flavor (mass) eigenstates respectively. The entries can be obtained at the leading order in $\cot\beta$:
\begin{align}
U_{11} &\sim U_{22} \sim U_{33}\sim 1 + \mO(\cot^2\beta) ,\nonumber\\
U_{12}& \sim -U_{21}\simeq\cot\beta , \nonumber \\
U_{13} &\sim -U_{31} \simeq\cot\beta \frac{v_2}{v_D} ,\nonumber\\
U_{23} & \sim - U_{32} \sim \mO(\cot\beta),
\label{eq:Umix}
\end{align}
where the expression of $U_{23}$ proceeds from a combination of terms proportional to  $\cot\beta$ and $\lambda_6$, and we set it as a free parameter. The more detailed expressions for the mass of CP-even Higgs and mixing matrix $U$ are given in the Appendix~\ref{sec:CPevenHiggs}. We can easily see that, in the decoupling limit, the modifications to SM Higgs  couplings with massive gauge bosons and the fermions arise at second order in $\cot\beta$, which are therefore at the percent level in our scenario since $\cot\beta \sim 0.1$.

\subsection{Fermion sector}
 Let's now turn to  mixing in the fermion sector, where we especially focus on the $b$ and $c$ quarks. As explained above,  the masses  of the $b$ and $c$ quarks come from the mixing with heavy vector like fermions $\psi_{b,c}$.  We first consider the mixing between $\psi_b$ and $b$. 
The $2\times 2$ mass matrix in $(\psi_b, b)$ basis simply  reads:
\begin{align}
M_b=\left(\begin{array}{cc}
m_{b,\psi} & \frac{ y_{3b,\psi} v_D}{\sqrt{2}} \\
\frac{ y_{2b,\psi} v_2}{\sqrt{2}} & 0
\end{array}\right)
\equiv \left(\begin{array}{cc}
m_{b,\psi} & m^b_{12} \\
m^b_{21} & 0
\end{array}\right), 
\end{align}
where we simply treat the off-diagonal terms as small variables $m_{12}^b \ll m_{b,\psi}$. It is straightforward to diagonalize the mass matrix by the orthogonal rotation of the left-handed and right-handed quark fields:
\begin{align}
\left(\begin{array}{c} \psi_{b,L} \\ b_L \end{array}\right)=\left(\begin{array}{cc}
c_{b,L} & s_{b,L} \\
-s_{b,L}   & c_{b,L} \end{array}
\right)\left(\begin{array}{c} \tilde{\psi}_{b,L} \\ \tilde{b}_L \end{array}\right) 
~~~~~~
\left(\begin{array}{c} \psi_{b,R} \\ b_R \end{array}\right)=\left(\begin{array}{cc}
c_{b,R} & s_{b,R} \\
-s_{b,R}  & c_{b,R} \end{array}
\right)\left(\begin{array}{c} \tilde{\psi}_{b,R} \\ \tilde{b}_R \end{array}\right),
\end{align}
where the mixing angles  are approximately given by:
\begin{align}
  s_{b,L}&\sim -\frac{m_{21}^b}{m_{b,\psi}},  \qquad s_{b,R}\sim-\frac{m_{12}^b}{m_{b,\psi}},
 \end{align}
and the mass eigenvalues are:
\begin{align}
m_{\tilde\psi_b}\simeq m_{b,\psi}  , \qquad m_{\tilde b} \simeq - \frac{m_{12}^{b} m^{b}_{21}}{m_{b,\psi}}\simeq -  s_{b,L}s_{b,R}m_{b,\psi},
\end{align}
where the mass formula for the bottom quark is similar to the partial compositeness scenario~\cite{Kaplan:1991dc}. The same analysis applies to the charm quark except the parameters are in the charm sector. The mass formula and the mixing angle are given by:
\begin{align}
m_{\tilde\psi_c}\simeq m_{c,\psi}  , \qquad m_{\tilde c} \simeq  - \frac{m_{12}^{c} m^{c}_{12}}{m_{c,\psi}} \simeq - s_{c,L}s_{c,R}m_{c,\psi}, \qquad 
  s_{c,L}\sim -\frac{m_{21}^c}{m_{c,\psi}}, \qquad   s_{c,R}\sim-\frac{m_{12}^c}{m_{b,\psi}}.
\end{align}

We now consider the mass eigenstates of $\chi_{1,2}$. The Dirac mass term for $\chi_2$ is simply:
\begin{align}
m_{\chi_2}=\frac{y_{\chi_2} v_D}{\sqrt{2}}
\end{align}
without any mixing with SM particles and this will be our dark matter candidate. At current stage, we assume the elastic DM scenario that Majorana mass $M_m = 0$, which can be originated from a global continuous symmetry for $\chi_2$. We will come back to Majorana DM later. There is  a mixing between $\chi_1$ and $\tau$ induced by the Dirac mass $m_{\tau1}$, which we assume to be tiny. So the mass eigenvalues at leading order are simply:
\begin{align}
m_{\tilde{\chi}_1} \simeq \frac{y_{\chi_1} v_D}{\sqrt{2}}, \qquad   m_{\tilde \tau} \simeq \frac{y_{\tau} v_2}{\sqrt{2}} 
\end{align}
At the linear order in $m_{\tau1}/m_{\chi_1}$, only the right-handed part  mix with each other:
\begin{align}
\left(\begin{array}{c} \chi_{1,R} \\ \tau_R \end{array}\right)=\left(\begin{array}{cc}
c_{\tau,R} & s_{\tau,R} \\
-s_{\tau,R}  & c_{\tau,R} \end{array}
\right)\left(\begin{array}{c} \tilde{\chi}_{1,R} \\ \tilde{\tau}_R \end{array}\right),
\end{align}
where the mixing angle are:
\beq
s_{\tau,R} \simeq -\frac{m_{\tau_1}}{m_{\tilde{\chi}_1}},  ~ s_{\tau, L} \simeq -\frac{m_{\tau_1}}{m_{\tilde{\chi}_1}} \frac{m_{\tilde{\tau}}}{m_{\tilde{\chi}_1}},
\eeq 
and we see clearly $s_{\tau, L} \ll s_{\tau, R}$ and can be neglected.

The relevance of  $s_{\tau,R}$ mixing  is to let the $\chi_1$ decay, so in principle we can make it as small as we want unless the lifetime of $\chi_1$ is long enough to have cosmological problems.  For  example, if we make it as small as $10^{-4}$, it will not affect the SM $\tau$ interactions in
any significant way and $\chi_1$ will have a decay width $\sim \alpha_{em} m_{\chi_1} s_{\tau, R}^2 \sim 10 \ \text{eV}$, implying that it will still decay promptly at the LHC.

\subsection{Gauge bosons interactions with fermions}
In this section, we will review the interactions between the fermions and the gauge bosons. 
Let us emphasize again that the gauge eigenstates of gauge bosons (e.g. $Z$ and $K$) are denoted without tildes, 
while the mass eigenstates (e.g. $\tilde{Z}$ and $\tilde{K}$) are denoted with tildes.
For the gauge bosons $W^\pm$ and photon $A$, no further mixing are induced by $U(1)_D$ and thus they
are the same as in SM.
First, we notice that in the gauge basis, the interaction Lagrangian in the quark sector reads:
\begin{eqnarray}
\mathcal{L}_{int}^I & = & \frac{g}{\sqrt{2}} W^+_\mu \bar{t}_L \gamma^\mu b_L
+  \frac{g}{2c_w}  Z_\mu\left( -\bar{b}_L \gamma^\mu b_L + \bar{c}_L \gamma^\mu c_L \right)  + g_D K_\mu \left( \bar{b}_R \gamma^\mu b_R - \bar{c}_R \gamma^\mu c_R  \right) \nonumber \\
&+& \frac{g s_w^2}{3c_w} Z_\mu \left(
\bar{b}\gamma^\mu b 
+\bar{\psi}_{b} \gamma^\mu \psi_{b} 
\right)
- \frac{2 g s_w^2}{3c_w} Z_\mu \left(
\bar{c}\gamma^\mu c 
+\bar{\psi}_{c} \gamma^\mu \psi_{c}
\right) ,
\end{eqnarray}
where we neglect the photon couplings as it is only determined by the electric charge of the fermions, not changing the couplings of $K$ and $Z$. To determine the couplings of $Z$, we separate the electric-charge ($Q$) part and the weak isospin part $T^3$.
Because the electromagnetic gauge symmetry is unbroken, only particles with the same electric-charge can mix with each other after EWSB, making the $Q$ part of the $Z$ couplings  flavor diagonal. Then the only flavor off-diagonal  $Z$ coupling comes from the $T^3$ contribution,
namely
\begin{align}
\frac{g}{2c_w}  Z_\mu\left( -\bar{b}_L \gamma^\mu b_L + \bar{c}_L \gamma^\mu c_L \right)
\end{align}
which are purely left-handed. In contrast, the $K$ couplings are purely right-handed. Now It is easy to obtain the gauge boson couplings in the mass mass eigenstate by performing the orthogonal rotation to the gauge bosons and the fermions. The results for the SM charge gauge bosons read:
\begin{align}
\mathcal{L}_{int}^{W} & =  \frac{g}{\sqrt{2}} W^+_\mu \bar{t}_L \gamma^\mu (c_{b,L} \tilde{b}_L 
- s_{b,L} \tilde{\psi}_{b,L})+ h.c.
\end{align}
and for the neutral $\tilde Z_\mu$ state the interactions read
\begin{align}
\mathcal{L}_{int}^Z & =
\tilde Z_\mu \left[  \bar {\tilde b}_L \gamma^\mu \tilde{b}_L\frac{g\cos\alpha}{c_w}\left(\frac{s^2_w}{3}-\frac{1}{2}c^2_{b,L} \right) + \bar{\tilde b}_R \gamma^\mu \tilde{b}_R\left(\frac{gs_w^2}{3c_w}\cos\alpha -g_D\sin\alpha c^2_{b,R}\right) \right]  \nonumber \\
&+ \tilde Z_\mu \left[  \bar {\tilde c}_L \gamma^\mu \tilde{c}_L\frac{g\cos\alpha}{c_w}\left(-\frac{2s^2_w}{3}+\frac{1}{2}c^2_{c,L} \right) + \bar{\tilde c}_R \gamma^\mu \tilde{c}_R\left(-\frac{2gs_w^2}{3c_w}\cos\alpha + g_D\sin\alpha c^2_{c,R}\right) \right]  \nonumber\\
&+ \frac{g \cos\alpha s_w^2 }{3c_w} \tilde Z_\mu \left\{
  \left[c_{b,L}^2\bar{\tilde{\psi}}_{b,L} \gamma^\mu \tilde\psi_{b,L} 
+\left( L \leftrightarrow R \right)\right] 
- 2 \left[c_{c,L}^2\bar{\tilde{\psi}}_{c,L} \gamma^\mu \tilde\psi_{c,L}+ \left( L \leftrightarrow R \right)
\right]
\right\}  \nonumber\\
&+  \tilde Z_\mu 
\left[
\left(  \frac{g \cos\alpha \ c_{b,L} s_{b,L}}{2 c_w}
\bar{ \tilde {\psi}}_{b,L} \gamma^\mu \tilde b_{L}   
+ g_D \sin\alpha \ c_{b,R} s_{b,R} \ \overline{ \tilde {\psi}}_{b,R} \gamma^\mu \tilde b_{R}  + h.c. \right) - \left(b \leftrightarrow c \right)   \right] 
\label{eq:ZbLagragian}
\end{align}
where the mixing angles are defined in the previous two sections. We can clearly see that the modifications to the $Z\bar{b}_Rb_R$ and the $Z \bar{c}_R c_R$ couplings come at linear order in $\sin\alpha$ and are of opposite sign, while for the left-handed couplings, they arise from the normalization of the quark fields starting at the square order of the mixing parameters $\sin^2\alpha, s_{c,L}^2, s_{b,L}^2$. As we will see later, a small modification to the left-handed bottom and charm $Z$ boson couplings is necessary in order to satisfy the total $b,c$ hadronic cross section  measurements on the $Z$-pole.

For the $U(1)_D$ gauge boson interactions at lowest order, we have:
\begin{align}
\mathcal{L}_{int}^K&=\frac{g\sin\alpha}{c_w} \tilde K_\mu J_{Z,q}^\mu + \tilde K_\mu \left[\bar {\tilde b}_L \gamma^\mu b_L \frac{g\sin\alpha}{c_w}\left(\frac{s^2_w}{3}-\frac{1}{2}c^2_{b,L} \right)+ \bar{\tilde b}_R \gamma^\mu \tilde{b}_R\left(\frac{gs_w^2}{3c_w}\sin\alpha + g_D \cos\alpha c^2_{b,R} \right) \right] \nonumber\\
& + \tilde K_\mu \left[\bar {\tilde c}_L \gamma^\mu c_L \frac{g\sin\alpha}{c_w}\left(-\frac{2s^2_w}{3}+\frac{1}{2}c^2_{c,L} \right)+ \bar{\tilde c}_R \gamma^\mu \tilde{c}_R\left(-\frac{2gs_w^2}{3c_w}\sin\alpha - g_D \cos\alpha c^2_{c,R} \right) \right]  \nonumber\\
&+\tilde K_\mu\left(\bar{ \tilde {\psi}}_{b,L} \gamma^\mu \tilde b_{L} \frac{g}{2c_w} \sin\alpha c_{b,L} s_{b,L} - \bar{ \tilde {\psi}}_{b,R} \gamma^\mu \tilde b_{R} g_D \cos\alpha c_{b,R} s_{b,R}  \right)   \nonumber \\
&-\tilde K_\mu\left(\bar{ \tilde {\psi}}_{c,L} \gamma^\mu \tilde c_{L} \frac{g}{2c_w} \sin\alpha c_{c,L} s_{c,L} -  \bar{ \tilde {\psi}}_{c,R} \gamma^\mu \tilde c_{R} g_D \cos\alpha c_{c,R} s_{c,R} \right) 
\label{eq:KbLagragian}
\end{align}
where  $J_{Z,q}^\mu$ is the SM quark neutral currents except the bottom and charm quarks:
\beq
J_{Z,q}^\mu = \sum_{q\neq b, c} (T^3 - Q s_w^2) \bar{q} \gamma^\mu q.
\eeq
We can see that $\tilde{K}_\mu$ mainly couples to  the SM right-handed bottom and charm quarks  with gauge coupling $g_D$ and couples universally to other quarks and leptons through its small mixing with $Z$ boson.
We finally comment that due to the existence of a Dirac mass  for the vector-like quark $\psi_b$ and $\psi_c$, one can
lift these vector-like fermion masses ($\gtrsim 1$ TeV) to decouple $\psi_b$ and $\psi_c$ from LHC physics, while choose appropriate mixing angles to give the right mass to the $b$ and $c$ quarks. 

Next we consider the gauge boson interactions in the lepton sector including $\tau$ and  $\chi_{1,2}$. The interaction Lagrangian  in gauge basis reads:
\begin{align}
\mathcal{L}_{\chi}& = -eA_\mu \left( \bar{\chi}_1\gamma^\mu \chi_1 + \bar{\tau}\gamma^\mu \tau\right)   +  Z_\mu \left( \frac{g s_w^2}{c_w} \left(\bar{\chi}_{1}\gamma^\mu \chi_{1} + 
 \bar{\tau}\gamma^\mu \tau \right) -
\frac{g }{2 c_w} \bar{\tau}_L\gamma^\mu \tau_L  \right)  \nonumber \\
& +K_\mu g_D \left( \bar{\chi}_{1,R}\gamma^\mu \chi_{1,R} - \bar{\chi}_{2,R}\gamma^\mu \chi_{2,R}  \right) \,.
\end{align}
In the mass eigenstate basis, the Lagrangian at leading order mixing is,
\begin{align}
\label{eq:gaugeboson-leptons}
\mathcal{L}_{\chi}& \simeq -eA_\mu \left( \bar{\tilde \chi}_1\gamma^\mu \tilde{\chi_1} 
+ \bar{\tilde{\tau}}\gamma^\mu \tilde{\tau} \right) + \frac{g\sin\alpha}{c_w} \tilde K_\mu J_{Z,\ell}^\mu \nonumber \\
&  + \tilde Z_\mu \left( \cos\alpha \left(\frac{gs_w^2}{c_w}(\bar{\tilde \chi}_1\gamma^\mu \tilde{\chi_1} + \bar{\tilde{\tau}}\gamma^\mu \tilde{\tau} )
 -\frac{g }{2 c_w} \bar{\tilde{\tau}}_L\gamma^\mu \tilde{\tau}_L  \right) 
 - g_D \sin\alpha \left(c_{\tau, R}^2 \bar{\tilde{\chi}}_{1,R}\gamma^\mu \tilde{\chi}_{1,R}
 - \bar{\chi}_{2,R}\gamma^\mu \chi_{2,R} \right) \right)  \nonumber \\
&+ \tilde K_\mu \left( \sin\alpha \left( \frac{gs_w^2}{c_w} (\bar{\tilde \chi}_1\gamma^\mu \tilde{\chi_1} + \bar{\tilde{\tau}}\gamma^\mu \tilde{\tau} )
 -\frac{g }{2 c_w} \bar{\tilde{\tau}}_L\gamma^\mu \tilde{\tau}_L  \right) 
 + g_D \cos\alpha \left(c_{\tau, R}^2 \bar{\tilde{\chi}}_{1,R}\gamma^\mu \tilde{\chi}_{1,R}
 - \bar{\chi}_{2,R}\gamma^\mu \chi_{2,R} \right) \right)   \nonumber\\
&+ g_D (\cos\alpha \tilde K_\mu -\sin\alpha \tilde Z_\mu) c_{\tau, R} s_{\tau,R} ( \bar{\tilde{\tilde{\chi}}}_{1,R} \gamma^\mu \tilde{\tau}_{R} +h.c.) .
\end{align}
where  $J_{Z,\ell}^\mu$ is the SM lepton neutral currents except the $\tau$:
\beq
J_{Z,\ell}^\mu = \sum_{\ell\neq \tau} (T^3 - Q s_w^2) \bar{\ell} \gamma^\mu \ell.
\eeq
As explained in previous subsection, $s_{\tau, R}$ can be chosen to be very small to make
$\chi_1$ decay promptly at LHC while not affecting the early cosmology. 
We note that $\chi_1$ has mass around $\sim v_D$, thus is relevant for LHC physics. Later we will show that due to its coupling only to hypercharge, it is not constrained by current LHC limits.

\subsection{Higgs interaction with Fermions and Gauge Bosons}
After we consider the mass eigenstates of Higgs and fermions, we can have the following interactions: 
\begin{align}
\mathcal{L}_{yuk}^I & =-\left(\frac{m_t}{v s_\beta}\bar{\tilde t}_L \tilde t_R + \frac{m_s}{v s_\beta}\bar{\tilde s}_L \tilde s_R  \right)(-ct_\beta H^0_1+ U_{22} H^0_2 +U_{23}  H^0_3)  \nonumber \\
&-\frac{m_{\tilde b}}{s_\beta v}\bar{\tilde b}_L \tilde {b}_R \left(-\left( ct_\beta c_{b,L}  + \frac{c_\beta s_\beta  v^2}{ v_D^2} c_{b,R}\right) H^0_1 +c_{b,L}U_{22}H^0_2 + \frac{s_\beta v}{v_D}c_{b,R}U_{33} H^0_3\right) \nonumber\\
&-\frac{m_{\tilde c}}{s_\beta v} \bar{\tilde c}_L \tilde {c}_R\left(-\left(ct_\beta c_{c,L}  + \frac{c_\beta s_\beta  v^2}{ v_D^2} c_{c,R}\right) H^0_1 + c_{c,L}U_{22}H^0_2 + \frac{s_\beta v}{v_D}c_{c,R}U_{33} H^0_3\right)  \nonumber\\
& - \left(\frac{m_{\chi_2}}{v_D}\bar{\chi}_{2,L} \chi_{2,R} + \frac{m_{\tilde{\chi}_1} c_{\tau,R}}{v_D}\bar{\tilde \chi}_{1,L}\tilde{\chi}_{1,R} + \frac{m_{\tilde{\chi}_1}s_{\tau,R}}{v_D}\bar{\tilde{\chi}}_{1,L}\tilde {\tau}_R \right)\left(-\frac{v}{v_D}c_\beta H^0_1 - U_{23} H^0_2+U_{33} H^0_3\right)  \nonumber\\
& -\left(\frac{m_{\tilde{\tau}} c_{\tau,R}}{s_\beta v}\bar{\tilde \tau}_{L}\tilde{\tau}_{R}-\frac{m_{\tilde{\tau}} s_{\tau,R}}{s_\beta v}\bar{\tilde{\tau}}_{L}\tilde {\chi}_{1,R}  \right) \left(-ct_\beta H^0_1 +U_{22} H^0_2 +U_{23}H^0_3\right),
\end{align}
where we have abbreviated $c_\beta \equiv\cos\beta, ct_\beta \equiv \cot\beta, \cdots etc$ and substituted the leading values for $U_{12}$ and $U_{13}$ in Eq.~(\ref{eq:Umix}). Note that we have only kept the leading term in the $H_2^0(H_3^0) b\bar{b}(c\bar{c})$ couplings in the limit $c_\beta \ll 1$. Since $s_\beta \simeq 1$, the SM-like Higgs boson $H_2^0$ will couple to SM fermions the same as Standard Model except from $\mO(c_\beta^2)$ corrections, which are at the percent level in our model. This implies that this model cannot be tested through Higgs fermion coupling measurements at the LHC and hence we shall not discuss these constraints anymore.  We also see that the $H_3^0$ is $\Phi_3$-like and coupled to  bottom and charm quark  proportional to their mass as discussed before. Note that it also couples to top quark through its mixing with $h_2^0$, which maybe relevant due to the large top Yukawa coupling and the mixing size of order $ct_\beta$. 

In the following, we  will  consider the mass hierarchy $m_{H_2^0}, m_{H_3^0} \lesssim m_{\tilde{\chi}_1} \ll m_{H_1^0} \ll m_{\tilde{\psi}_{b,c}}$.  Hence, the heavy charged  lepton $\chi_1$ can decay to scalars plus $\tau$ leptons, where the leading channel is $\tau H_3^0$ which is only suppressed by $s_{\tau,R}^2$, while the channel $\tau H_2^0$ is further suppressed by tau mass.
Given Eq.~(\ref{eq:gaugeboson-leptons}), the other dominant decay channel for $\chi_1$ is $\tau \tilde{K}$ which is also of order $s_{\tau,R}^2$. Therefore, $\tilde{\chi}_1$ decays into $\tau (\bar{b}b)$ and $\tau (\bar{c}c)$, which could be a new signature to look for at LHC depending on the production cross section of $\chi_1$.

For completeness, we list the leading interaction between $\psi$ and $c, b$, and neglect the quadratic terms like $\mathcal{O}(s_{b,c}^2, s_{b,c} c_\beta)$,
\begin{align}
-\mathcal{L}_{yuk}^I & \supset s_{b, L} m_{\tilde{\psi}_b} \frac{U_{22} H_2^0}{v_2} c_{b, L} c_{b, R} \bar{\tilde{b}}_L \tilde{\psi}_{b, R} + s_{b, R} m_{\tilde{\psi}_b} \frac{U_{33} H_3^0}{v_D} c_{b, L} c_{b, R} \bar{\tilde{\psi}}_{b, L} \tilde{b}_{ R} + h.c.  
\nonumber \\
& + s_{c, L} m_{\tilde{\psi}_c} \frac{U_{22} H_2^0}{v_2} c_{c, L} c_{c, R} \bar{\tilde{c}}_L \tilde{\psi}_{c, R} + s_{c, R} m_{\tilde{\psi}_c} \frac{U_{33} H_3^0}{v_D} c_{c, L} c_{c, R} \bar{\tilde{\psi}}_{c, L} \tilde{c}_{ R} + h.c. \ .
\end{align}
Note that the couplings to diagonal heavy quark $\bar{\tilde{\psi}} \tilde{\psi}$ are neglected at $\mathcal{O}(s_{b,c}^2)$. The vector-like quark $\tilde \psi_{b,c}$ can decay into $\tilde b, \tilde c$ quarks plus $\tilde Z$, $\tilde K$ and scalars. The decay width to $\tilde{Z}, \tilde{K}$, $H_2^0, H_3^0$ are proportional to $s_{q, L}^2$, $s_{q, R}^2$, 
$s_{q, L}^2 m_{\tilde{\psi}_q}^2/v_2^2$, $s_{q, R}^2 m_{\tilde{\psi}_q}^2/v_D^2$. Given that the Dirac mass of $\psi$ is much larger than $v_2 \sim v_D$, the dominant decay channels for $\psi_{b, c}$ are $b,c$ plus scalars. Since one can give a large enough Dirac mass for $\tilde{\psi}_{b,c}$ to evade the collider constraints, we will not further discuss their search at LHC. 

Next, we consider the Yukawa interaction with charged Higgs $H^\pm$ and CP odd Higgs $A^0$. The Lagrangian  for the charged Higgs in the mass eigenstates reads: 
\begin{align}
\mathcal{L}^{H^\pm}_{\rm int}&= 
+ \frac{\sqrt{2} m_b }{t_\beta v} \bar{t}_L H^+ \left( \tilde{b}_R + \frac{c_{b,R}}{s_{b,R}} \tilde{\psi}_{b,R} \right)   
-\frac{\sqrt{2} m_t}{t_\beta v} \left(  c_{b,L} \bar{\tilde{b}}_L - s_{b,L} \bar{\tilde{\psi}}_{b,L}\right)  H^- t_R 
+ h.c.  \nonumber \\
& 
+\frac{\sqrt{2} m_s}{t_\beta v} \left(c_{c,L} \bar{\tilde{c}}_L - s_{c,L} \bar{\tilde{\psi}}_{c,L} \right) H^+ s_R
  -\frac{\sqrt{2}m_c}{t_\beta v} \bar{s}_L H^- \left(\tilde{c}_R + \frac{c_{c,R}}{s_{c,R}} \tilde{\psi}_{c,R} \right)
  +h.c.  ,
 \label{eq:HpmInt}
\end{align}
The fermion interaction with $A^0$ is given in the Appendix~\ref{sec:CPoddHiggs}. As discussed before, $H^\pm$ and $A^0$ can be made as heavy as TeV, therefore we are not going to discuss them further.

We finally list the interactions between one CP-even scalar and two gauge bosons, which maybe relevant for the LHC phenomenology. The Lagrangian in the gauge basis at leading $c_\beta$ order is :
\beq
\begin{split}
\mL_{\phi VV} &= \frac{2m_W^2}{v} W^{+\mu} W^-_{\mu} \left(c_\beta h_1^0 + s_\beta h_2^0 \right)  + \frac{m_Z^2}{v}   \left[c_\beta  h_1^0\left(Z_\mu + \frac{2 g_D c_w}{g} K_\mu\right)^2 + s_\beta h_2^0  Z_\mu Z^\mu\right]  + h_3^0  \frac{m_K^2}{v_D} K_\mu^2 ,
\end{split}
\eeq
where the couplings of gauge bosons with the scalars are determined by the scalars' contributions to the mass of the gauge bosons. The Lagrangian for the mass eigenstates are:
\beq
\begin{split}
\mL_{\phi VV} 
&\simeq \frac{2 m_W^2}{v} W^{\mu +}W_\mu^- \left( c_\beta\left(U_{11}-1\right) H^0_1 +s_\beta U_{22} H^0_2 +s_\beta  U_{23} H^0_3\right) \\ 
&+ \frac{m_Z^2}{v}\tilde Z^\mu \tilde Z_\mu \left( c_\beta \left( \left(U_{11}-1\right) c_\alpha^2  +s_{2\alpha}\frac{ g_D v}{m_Z} \right )H^0_1 + c^2_\alpha s_\beta U_{22} H^0_2 + c^2_\alpha  s_\beta U_{23}H^0_3\right) \\ 
&+\tilde K_\mu \tilde K^\mu \left(\left(c^2_\alpha c_\beta g_D^2 v (U_{11}- 1) -c_\beta s_{2\alpha} g_D m_ZU_{11}\right) H^0_1 -c^2_\alpha \frac{m_K^2}{v_D} U_{23} H^0_2  +c^2_\alpha \frac{m_K^2}{v_D}U_{33}H^0_3 \right) \\
&+\tilde Z_\mu \tilde K^\mu \left(-2c^2_\alpha c_\beta g_D m_Z U_{11} H^0_1 + \frac{s_{2\alpha} s_\beta m_Z^2 }{v} U_{22}H^0_2 -s_{2\alpha}\frac{m_K^2}{v_D}U_{33} H^0_3\right) ,
\end{split}
\eeq
where we have kept leading terms in $c_\beta$ and $s_\alpha$ for $H^0_{1,2,3}$ term respectively. We can see that  $H_2^0$ couplings to gauge bosons are modified at the percent level $ \sim c_\beta^2$, which is consistent with the present  precision at the LHC. The $H_1^0$ couplings are further suppressed at quadratic or cubed order, $\mathcal{O}(c_\beta^3, c_\beta s_\alpha, s_\alpha^2 )$, though linearly suppressed by $c_\beta$ for $\tilde{Z}\tilde{K}$ coupling, while $H_2^0, H_3^0$ are at most suppressed by linear $c_\beta$ or $s_\alpha$. This fact reveals that it is much more difficult to search for $H^0_1$ at the LHC. 

For the $H_3^0$, it couples largely to the $\tilde{K}$ gauge boson as it is the main source of $\tilde{K}$ gauge boson mass. As a result,  if $m_{\chi_1,\chi_2}> m_{H^0_3}/2$, it will dominantly decay into $\tilde{K}$ pair if this decay channel is kinematically open. It can also decay into $\tilde b\bar{\tilde b}$, $\tilde c\bar{\tilde c}$ pairs which may be dominant if the $\tilde{K}$ decay channel is closed. It could decay into other SM fermions pair but will be suppressed by the mixing between $H^0_2$ and $H^0_3$.  Concerning its production at the LHC, we expect that it is mainly produced through $gg$ fusion due to top  and bottom loops. If $U_{23} $ is of order $ct_\beta$,  top loop will dominate. In this case, its production cross section at the LHC will be suppressed by $ct_\beta^2 \simeq 0.01$ compared with a SM-like Higgs boson of the same mass, namely  around $\sigma_{13\text{TeV}}(pp \to H^0_3) \simeq 0.44 $~pb and $\sigma_{13\text{TeV}}(pp \to H^0_3 j j) \simeq  $~0.037pb for  $m_{H_3^0} = 125 \GeV$. These cross sections are too small to discriminate the $H^0_3$ production from the multi-jet QCD background. If $m_{\chi_2}< m_{H^0_3}/2$, the most promising scenario for searching $H^0_3$ is $H^0_3 jj$ production, following by the nearly $100\%$ invisible decay to $\chi_2\bar\chi_2$, if $m_{\chi_1} , m_{\tilde K} > m_{H^0_3}/2$. Comparing to the cross section of $\sigma (jj (Z\to \nu\bar{\nu}))\sim 10^3$ pb, $H_3^0$ production is still hard  to probe at the LHC.

\section{Electroweak Precision Measurements}
\label{sec:EWPT}
\begin{figure}
  \centering
  \includegraphics[width=0.45 \columnwidth]{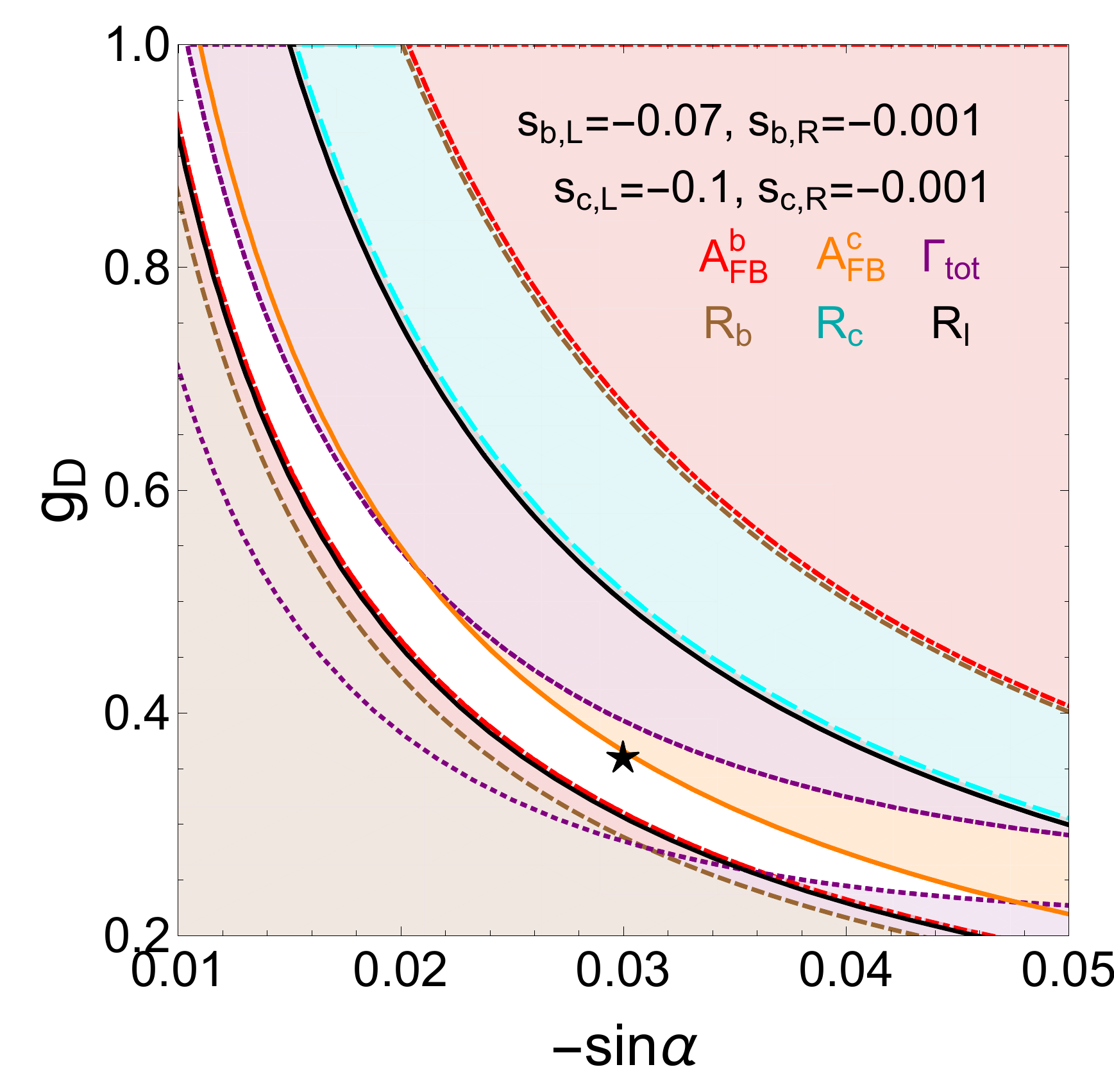} 
   \includegraphics[width=0.45 \columnwidth]{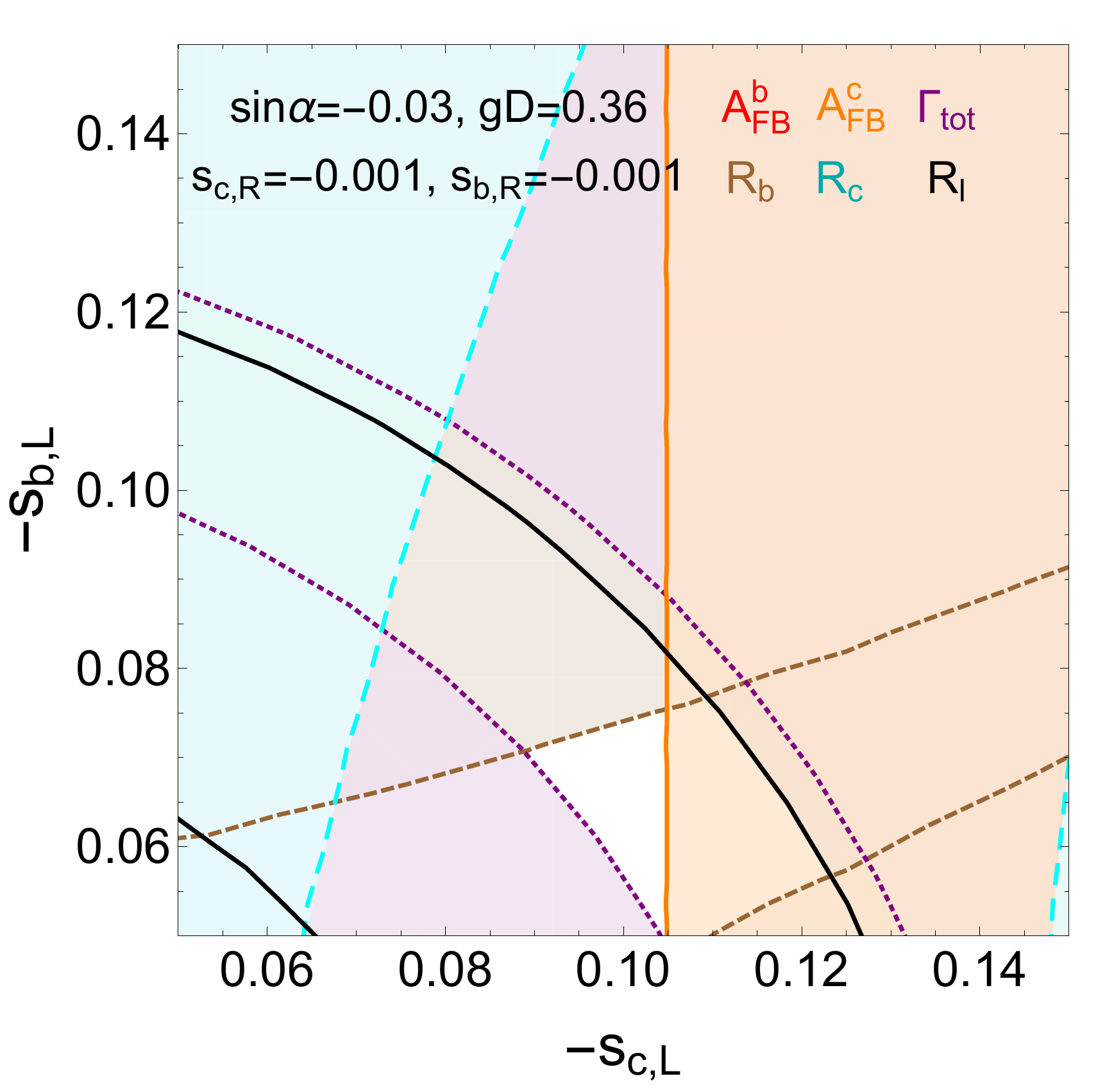} 
  \caption{ 
The color lines represent the $1\sigma$ bounds on different Z pole precision observables $R_b$, $R_c$, $R_L$, $A_{FB}^b$, $A_{FB}^c$ and $\Gamma_{tot}$.  The color shaded regions are excluded by these measurements at the $1\sigma$ level.  The white areas are the allowed regions by those precision measurements at $1\sigma$.
The black star in the left panel is our benchmark point.   }
  \label{fig:Rb-AFb-constraint}
\end{figure}

The main motivation behind this model is the observed 3~$\sigma$ deviation of the bottom-quark forward-backward asymmetry $A^b_{\rm FB}$ measured at the LEP experiment at CERN. It is well known that this asymmetry may be modified by varying the right-handed bottom coupling to the Z-boson~\cite{Choudhury:2001hs,Agashe:2006at,Luo:2007me,Djouadi:2006rk,Berger:2009hi,DaRold:2010as,Gori:2015nqa}. 
In general, the modification of the couplings produces other effects that 
have relevant implications on the precision electroweak observables, which should be considered simultaneously. In fact, the 
strongest constraints on this model come precisely from the Electroweak precision measurements~\cite{ALEPH:2010aa,Group:2009ae,Alcaraz:2006mx,ALEPH:2005ab,LEP:2003aa} including the T parameter and the $Z$-pole observables. In our setup, the mixing between $\tilde K$ and $\tilde Z$ will induce the custodial symmetry breaking, which modifies the $\tilde Z$ mass without changing the mass of the $W$ boson. The corresponding contribution to the T-parameter is given by:
\beq
 \hat{\alpha}(m_Z) T = - \frac{\Delta  m_Z^2}{ m_Z^2} \sim  \sin^2\alpha \frac{m_K^2 - m_Z^2}{m_Z^2} , \qquad  
\eeq 
where $\hat{\alpha}(m_Z) $ is the value of the fine-structure constant evaluated on the $Z$-pole, whose value is \cite{Jegerlehner:2009ry,Hanneke:2008tm}:
\beq
 \hat{\alpha}(m_Z)  = \frac{1}{127.95}.
 \eeq
The modification of the T-parameter has the same sign as $m_K^2 - m_Z^2$. From the T-parameter measurement  $T = 0.08 \pm 0.12$ \cite{Agashe:2014kda}, we can obtain the $95\%$ bound on the modification of the $Z$ mass:

\beq
  \sin^2\alpha \frac{m_K^2 - m_Z^2}{m_Z^2} \in [-0.00121,0.00246]
\eeq
which can translated into the bound on the mixing angle $\sin\alpha$ for given mass of the $\tilde K$ gauge boson.

Next, we consider the $Z$-pole measurements, including not only $A^b_{\rm FB}$ but also the total width of the $Z$ boson  $\Gamma_{\rm tot}$, the  heavy flavor quarks (bottom and charm quark) production ratio $R_{b,c}$, lepton production ratio $R_l$, and the forward-backward asymmetry of the charm quarks $A_{FB}^{c}$. They can be roughly written in terms of the left-handed and right-handed $Z$-couplings as:
\begin{align}
R_{b,c}&\equiv\frac{\Gamma(\tilde Z\to \tilde b\bar{\tilde b}(\tilde c\bar{\tilde c}))}{\Gamma(\tilde Z\to {\rm hadrons})}\simeq \frac{\left(g_L^{(\tilde b,\tilde c)}\right)^2+\left(g_R^{(\tilde b,\tilde c)}\right)^2}{\sum_q \left(g_L^q\right)^2+\left(g_R^q\right)^2} \\ \nonumber
A_{FB}^{b,c}&= \frac{3}{4}\mA_{e}\mA_{b,c}\simeq \frac{3}{4}\mA_{e}\frac{\left(g_L^{(\tilde b,\tilde c)}\right)^2-\left(g_R^{(\tilde b,\tilde c)}\right)^2}{\left(g_L^{(\tilde b,\tilde c)}\right)^2+\left(g_R^{(\tilde b,\tilde c)}\right)^2} \\ \nonumber
R_{\ell}&\equiv\frac{\Gamma_{\rm{hadron}}}{\Gamma_{\ell\ell}} 
\end{align}
where we have neglected the masses of SM quarks and leptons.  We defined the  coupling ratio factor:
\begin{align}
 \mA_f\equiv\frac{\left(g_L^{f}\right)^2-\left(g_R^{f}\right)^2}{\left(g_L^{f}\right)^2+\left(g_R^{f}\right)^2}
\end{align}
for any of the SM quarks and leptons.
The $\tilde Z$ coupling expressions in Eq.~(\ref{eq:ZbLagragian}) has been used. In particular, the coupling between $\tilde Z$ and $\tilde b$ is changed due to the mixing between $\tilde K$ and $\tilde Z$
\beq
\delta g_{\tilde Z \tilde b_R {\bar \tilde b}_R}\sim -  g_D\sin\alpha~ c^2_{b(c),R} .
\eeq
Note that  the values of the mixing angles for the bottom and charm quarks with the heavy vector-like quark are constrained by the requirement of correctly reproducing the bottom and charm mass:
\beq
|s_{c,L} s_{c,R}| \sim \frac{m_c}{m_{c,\psi}}  \lesssim 5 \times 10^{-4} , \qquad |s_{b,L} s_{b,R}| \sim \frac{m_b}{m_{b,\psi}}  <  2.7 \times 10^{-3},
\label{eq:bcmassreq}
\eeq
where we have required the masses of heavy vector-like quarks to be larger than 1 TeV to satisfy the LHC direct search bounds, and the running mass of the bottom and charm quark at the 1 TeV scale has been used. This makes all mixing angles naturally small and hence the $c_{b,(c),R} \simeq 1$.

In  Fig.~\ref{fig:Rb-AFb-constraint}, we present the 1$\sigma$ bounds on the different precision measurements, considering the measurement of $A^{b, c}_{\rm FB}$, $R_{b,c,l}$ and $\Gamma_{\rm tot}$.  The constraints coming from
different measurements are represented by different colors, and the shaded areas are excluded at the $1\sigma$ level, with colors corresponding
to a superposition of the colors associated to the observables that lead to a constraint in that region of parameters. Most importantly, the white
bands are allowed by all precision measurements at the $1\sigma$ level and can fit the deviation of the forward-backward asymmetry $A_{\text{FB}}^b$ within $1\sigma$.
Combing all the electroweak precision measurements and T parameter constraint, we find out the preferred parameter space of $g_D$ and $\sin\alpha$ is 
\begin{align}
g_D\sin\alpha\sim -0.011.
\end{align}  
And we also fix the other mixing angels
\begin{align}
s_{b,L} = -0.07, s_{b,R}=-0.001, s_{c,L}=-0.1, s_{c,R}=-0.001.
\end{align}
Note that the observables $s_{c(b),R}$ have much weaker impact on the electroweak precision measurement compared with $s_{c(b),L}$, because 
the $\tilde{Z}$ coupling to SM fermion in Eq.~(\ref{eq:ZbLagragian}) contains $s_{c(b),R}$ only from $\tilde{Z}$ and $\tilde{K}$ mixing. The other change in the coupling  come from the left-handed mixing angles $c_{c(b),L}$.

Considering a benchmark point for which $m_{\tilde K}=115$ GeV, the constraint from the measurement of the $T$ parameter requires  $|\sin\alpha|< 0.064$.  Recall the modification of the $Z b_R \bar{b}_R$ coupling in Eq.~(\ref{eq:ZbLagragian}) neglecting the tiny bottom mixing angles:
\beq
\delta g_{\tilde Z \tilde b_R {\bar {\tilde b}}_R}\sim -  g_D\sin\alpha 
\eeq
The T parameter constraint  can also been rewritten as:
\beq
\frac{(\delta g_{\tilde Z \tilde b_R {\bar {\tilde b}}_R})^2}{g_D^2}\frac{m_K^2 - m_Z^2}{m_Z^2} \in [-0.00121,0.00246]
\eeq
This clearly put a bound in the $m_{\tilde K}-g_D$ plane for fixed value of $\delta g_{\tilde Z \tilde b_R {\bar {\tilde b}}_R}$, which is shown as orange region in Fig.~\ref{fig:Constraint-Kd} for $\delta g_{\tilde Z \tilde b_R {\bar {\tilde b}}_R} =  0.011$.  Note such value can solve the $A_{FB}^b$ discrepancy. We can see clearly that the
constraints on the T parameter  almost  exclude the lower half of the parameter space. Since we will take $g_D, m_{\tilde K}, \sin\alpha$ as input parameters, the $c_{\beta}^2$ can be written as:
\begin{align}
c_\beta^2  \sim-  \frac{\Delta m_Z^2}{m_Z^2}\frac{\sqrt{g^2 + g'^2}}{2\delta g_{\tilde Z \tilde b_R {\bar {\tilde b}}_R}},
\label{eq:sinalpha}
\end{align}
where we can easily see that in order to  modify $A^b_{\rm FB}$ at the desired value and be consistent with T parameter constraint, we need  $c_{\beta}^2 \lesssim  0.08$. It indicates the vev of $\Phi_1$ should be small, i.e.
$v_1 \lesssim 75$ GeV.

\section{$\tilde K$ searches at colliders }
\label{sec:K-Production}

\begin{figure}
	\includegraphics[width=0.46 \columnwidth]{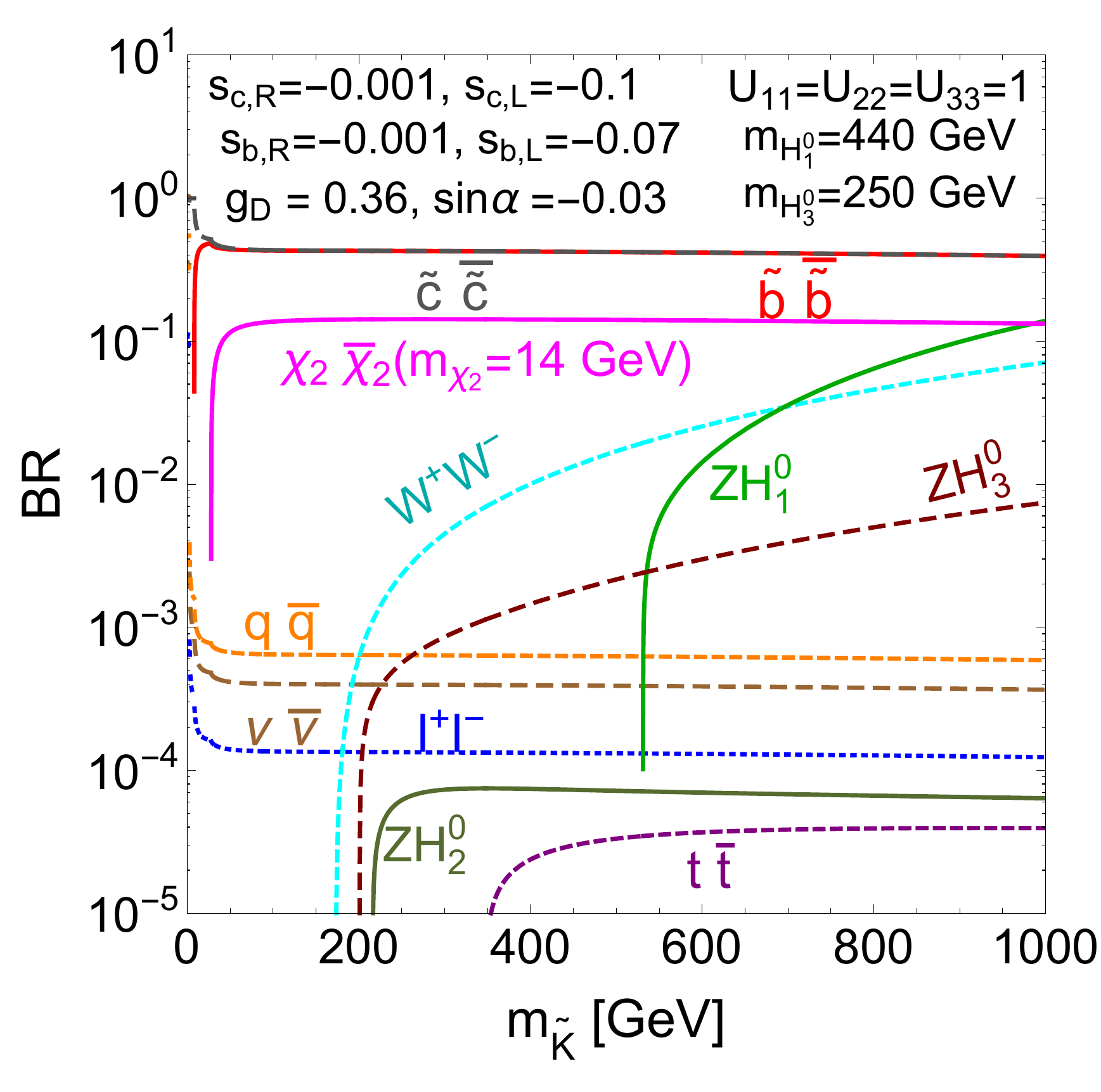} 
	\includegraphics[width=0.46 \columnwidth]{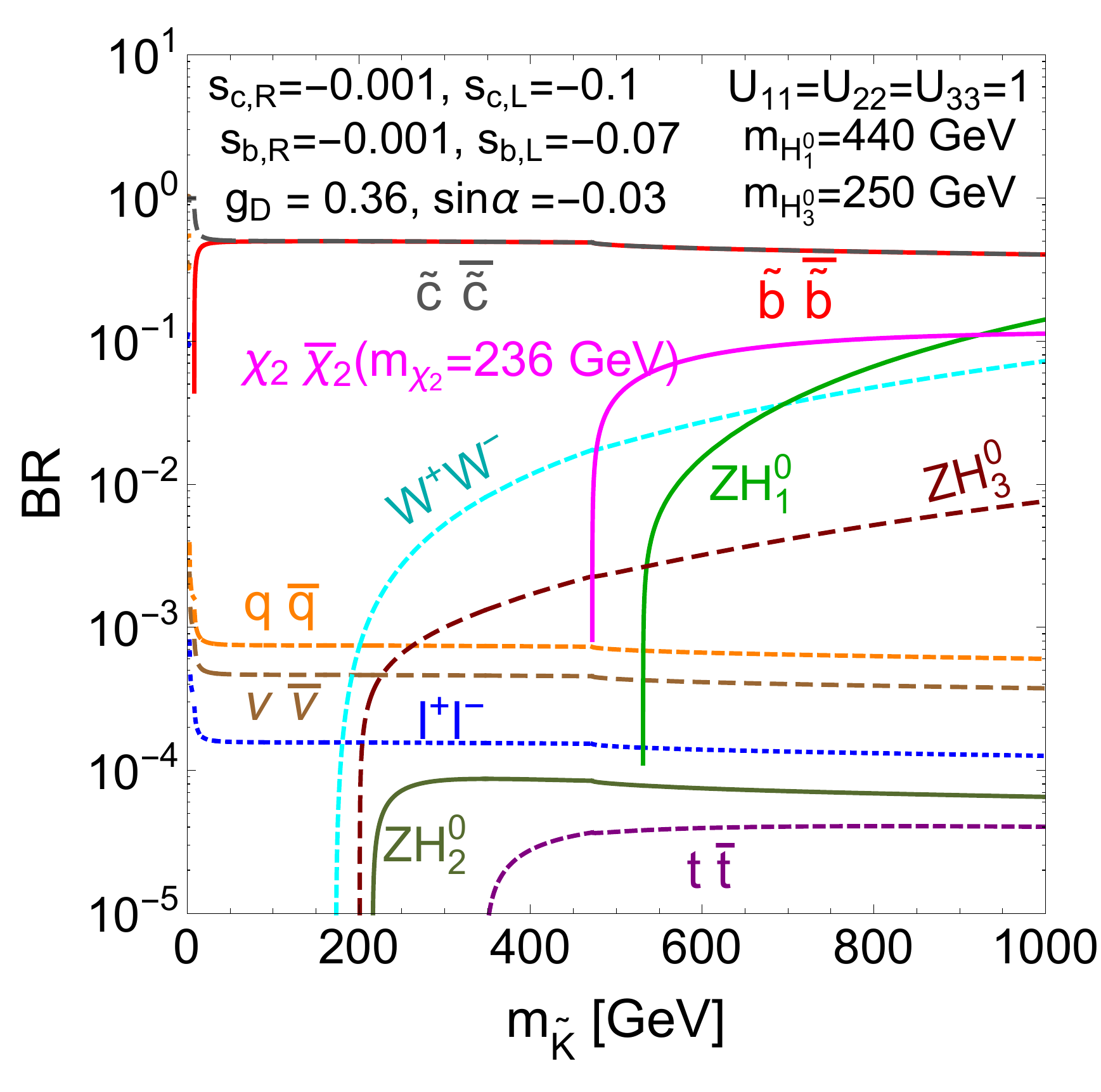} 
	\caption{The decay branching ratio for $\tilde K$. Left panel is for low mass $\chi_2$ and right panel is for high mass $\chi_2$,
		which are two DM benchmarks in Sec.~\ref{sec:DM-search}.
	}
	\label{fig:BR-Kd}
\end{figure}

In this section, we will consider the phenomenology of $\tilde K$ at the LHC.  Since our $\tilde K$ only coupled with  bottom and charm quarks before the small mixing between the $\tilde Z$ boson, its main production channel will be $\tilde b\bar{\tilde b}$ and $\tilde c\bar{\tilde c}$ initiated processes. It will also mainly decay into bottom and charm quarks with roughly  the same branching ratio $\sim 50\%$. The decay into leptons will be highly suppressed by the small mixing. We present the decay branching ratios of $\tilde{K}$ in Fig.~\ref{fig:BR-Kd}. There could be another decay channel of $\tilde K \to \chi_2\bar{\chi}_2$ if $m_{\chi_2} < m_{\tilde K}/2$, which would be around $1/7$  due to the color factor counting in low mass limit.

\begin{figure}
	\includegraphics[width=0.6 \columnwidth]{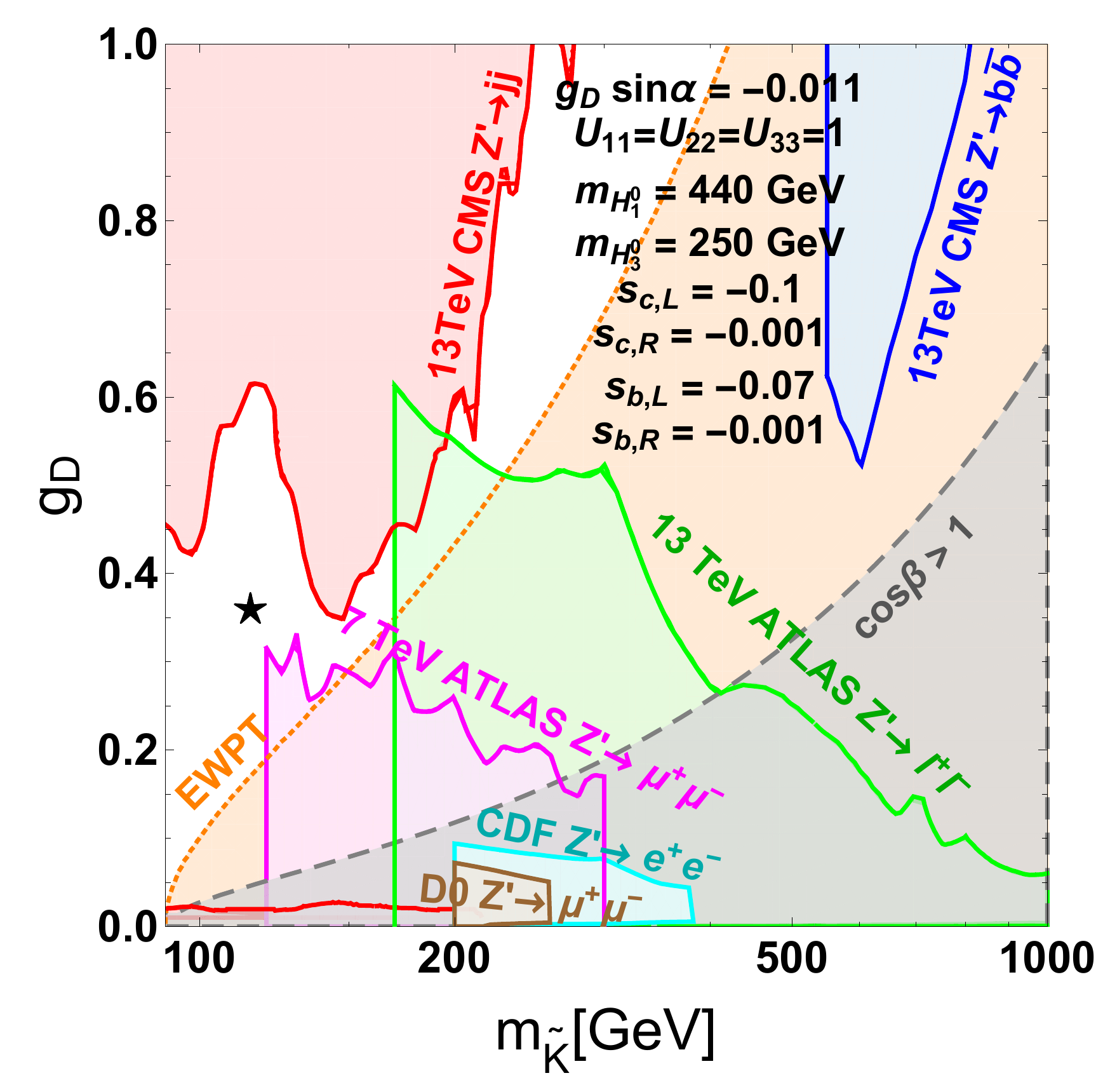}
	\caption{ The constraints from collider searches on $\tilde K$ for $g_D \sin\alpha = -0.011$. The mixing angle between heavy vector-like quarks and SM $b,c$ quarks are chosen following Fig.~\ref{fig:Rb-AFb-constraint}, where the EWPT constraint is the T parameter constraint under such choice. The red (blue) shaded regions correspond to  exotic $Z'$ search in dijet (b-jet pair) channel from CMS at 13 TeV~\cite{CMS:2017dhi} (\cite{CMS:2016ncz}), labeled as ``13 TeV CMS $Z' \to jj$" (``13 TeV CMS $Z' \to \bar{b}b$"). The $Z' \to \ell^+ \ell^-$ constraints from D0 and CDF are shown as brown and cyan area, labeled as ``D0 $Z' \to \mu^+ \mu^-$" and ``CDF $Z' \to e^+ e^-$". The $Z'$ dilepton searches at LHC are shown as magenta and green shaded area, from 7TeV \cite{Aad:2012cfr} and 13TeV \cite{Aaboud:2017buh} ATLAS, labeled as ``7TeV ATLAS $Z'\to \mu^+\mu^-$" and ``13TeV ATLAS $Z'\to \ell^+\ell^-$". The gray region is excluded because $\cos\beta >1$, while above the line has $\cos \beta<1$.
}
	\label{fig:Constraint-Kd}
\end{figure}

The presence of the light gauge boson $\tilde K$ is subject to several constraints. 
The first constraint comes from the exotic  $Z'$ decaying to dijet which associated produced with a jet from CMS~\cite{CMS:2017dhi} at 13 TeV, which is shown as red region in Fig.~\ref{fig:Constraint-Kd} . We see that there is a deep valley around 115 GeV, which is associated
with an interesting $2.9 \sigma$ local excess in that region of invariant masses.
CMS and ATLAS also  search for exotic $Z'$ decay to b-jet pair \cite{ATLAS:2016fol, CMS:2016ncz} at 13 TeV, but focus on the mass region around $550- 1500$ GeV. We only show the constraint from CMS as the blue region in Fig.~\ref{fig:Constraint-Kd} since CMS  present the constraint on the cross-section directly. The parameter spaces considered by ATLAS and CMS are not relevant to our analysis since they were already excluded by the $T$-parameter constraints, when  the $A_{FB}^b$ anomaly is considered by requiring $g_D \sin\alpha = -0.011$.

The next constraint is the exotic $Z'$ leptonic decay. We consider all the corresponding LHC searches at 7 TeV~\cite{Aad:2011xp}, 8 TeV~\cite{Aad:2014cka}, 13TeV~\cite{Aaboud:2017buh,ATLAS:2016jcu} and also the
Tevatron 1.96TeV searches by  D0 and CDF~\cite{RyanHooper:2004xma,Aaltonen:2007al}. Comparing all the searches, the most stringent constraint comes from the  13TeV ATLAS search \cite{Aaboud:2017buh} (green shaded) which goes down to 170 GeV. The constraints from D0 and CDF are shown as brown and cyan area.  We also show T parameter constraint in Fig.~\ref{fig:Constraint-Kd}  as orange area. 

LHC also searched for the low mass scalar in its leptonic decay. For our benchmark point, the branching ratio of  $\tilde {K}$ to $e^+e^-$, $\mu^+\mu^-$ and $\tau^+\tau^-$ are the same, which is $8.7\times 10^{-5}$. The most recently research is done by ALTAS \cite{Aad:2012cfr} at 7TeV and the constraint is $\sim 0.1$pb around mass $m_{\phi} =120$GeV, which is the lowest mass they considered in the $\mu^+\mu^-$ channel. For our benchmark point, the cross section for $pp\to ( \tilde K\to \mu^+\mu^-) = 0.08$pb at $m_{\tilde{K}} =115$GeV, which is again marginal within the constraint from ATLAS. The constraint  is shown as the magenta shaded area of Fig.~\ref{fig:Constraint-Kd}. 

Before closing this section, we comment on the intriguing hints of lepton flavor non-universality observed in the $R_K^{(*)}$ \cite{Aaij:2014ora, Aaij:2017vbb}  processes at the LHCb experiment and also in $R_D^{(*)}$  processes at the BaBar experiment~\cite{Lees:2012xj, Lees:2013uzd} and at LHCb~\cite{Aaij:2015yra} in charged lepton decay channel with tau leptons, though only weakly supported by Belle \cite{Huschle:2015rga, Sato:2016svk} and the recent LHCb result~\cite{Aaij:2017deq} from three-prong tau lepton decays. 

In our model, the gauge boson
$K$ couples flavor diagonally to b and c quark and hence not in a flavor universal way, which is similar to Ref.~\cite{Kamenik:2017tnu, Sala:2017ihs}. In this case, the $W^{\pm}$ loop 
effects can introduce
flavor changing coupling between the $K$ boson and $b$, $s$ quarks. However, the leptons
couple with $K$ only via $Z$ boson mixing, and hence the gauge boson couplings are lepton flavor universal. Therefore, our model
is unlikely to address $R_K^{(*)}$, unless we introduce, for example, muon leptons
charged under $U(1)_D$. 
Thus, it needs further considerations to reconcile 
$R_K$ or $R_K^*$ problems with bottom quark forward-backward asymmetry problem, 
what is beyond the scope of this paper. 
For $R_D^{(*)}$ lepton flavor non-universality, the charged
Higgs extension in type-II 2HDM has been excluded by the combination of $R_D$ and $R_D^{*}$
\cite{Lees:2012xj}. In our model, the $U(1)_D$ assignment of $\Phi_1$ determines that it is similar to
type-I 2HDM. In this case, the charged Higgs coupling to quarks are 
suppressed by $\cot \beta$, which we take to be small, and its contributions would be further suppressed by 
the fact that the masses of our Higgs bosons $H_0, A_0 $ and $H^\pm$ are large, of order of a few to several TeV,  which further reduces
their relevance to $R(D^*)$. 
Actually one might try changing $U(1)_D$ charge of $\Phi_1$
from $X$ to $-X$ in order to write down SM Yukawa coupling for $\Phi_1$.
However, this  induces the wrong sign for $\sin \alpha$, which forces us to stay
with the current charge assignment in Tab.~\ref{tab:gaugecharge}. Therefore, we conclude that
an extension of this model would be necessary to solve the flavor problems in $R_K^{(*)}$ and $R_D^{(*)}$
together with the bottom-quark forward-backward  asymmetry. We shall not explore such an extension in this article. 

\subsection{Benchmark for bottom-quark forward-backward asymmetry}

Based on the constraints from electroweak precision measurement and $\tilde K$ search at LHC, we set our benchmark point as $m_{\tilde K} = 115$ GeV, $g_D =0.36$ and $\sin\alpha= -0.03$, resulting $\tan\beta=7.4$. The choice of $g_D$ and $\sin \alpha$ satisfies the constraints from Z-pole observables in Fig.~\ref{fig:Rb-AFb-constraint}, which especially can also lead to $1\sigma$ agreement for the bottom-quark forward-backward asymmetry. In Fig.~\ref{fig:Constraint-Kd}, the collider limits of $\tilde{K}$ and T parameter still allow its mass to be around $\left[100, 140 \right] $. We do not consider degenerate masses between $Z$ and $K$ which may leads to large mixing. $m_{\tilde K} = 115$ is chosen because of the interesting $2.9\sigma$ local excess in Ref.~\cite{CMS:2017dhi}, but other $m_{\tilde{K}}$ around this region is also plausible. 

Note that the mass of the new gauge boson $\tilde K$ is very close to SM Higgs mass. For this benchmark point, the Drell-Yan cross section for $\tilde K$ production at the 13 TeV LHC will be sizable, around $3.1\times 10^3$ pb. The associated production cross section at LHC with another one or two jets are also listed in Table~\ref{tab:benm}.

\begin{table}[!htb]
\begin{tabular}{|c|c|c|c|c|c|c|c|}
\hline
 $g_D = 0.36,  \sin\alpha = -0.03$  & 7TeV & 8TeV&13TeV \\
    \hline
$\sigma(pp \rightarrow \tilde K)$  [pb] & $1.0\times 10^3 $ & $1.3\times 10^3$ & $3.1\times 10^3$  \\
\hline
$\sigma(pp \rightarrow \tilde K j)$ [pb] & $3.6\times 10^2$ & $4.8\times 10^2$& $1.3\times 10^3$ \\
\hline
$\sigma(pp \rightarrow \tilde K jj)$  [pb]  & $1.3\times 10^2 $ & $1.8\times 10^2 $ & $6.7\times 10^2$  \\
\hline
\end{tabular}
\caption{The cross sections for $\tilde K$ production in Drell-Yan channel and jet associated channels. For the cross section calculation, the mixing between $\tilde K$ and $ \tilde Z$ is not included due to small value of $\sin \alpha$.}
\label{tab:benm}
\end{table}

For our benchmark point $m_{\tilde K} = 115$ GeV, since it can decay into $\tilde b\bar{\tilde b}$ at around $50\%$, it can easily fake a $b\bar{b}$ decaying SM Higgs boson $m_h$ = 125 GeV at the LHC, because the large uncertainty for reconstructing hadronically decaying particles. In this case, it is important to check the constraints coming from SM Higgs searches with the Higgs decaying into bottom quark pairs. To calculate the cross-sections
in our model, we have used FeynRules 2.0~\cite{Alloul:2013bka} to generate the model files and implement it in 
MadGraph5$\textunderscore$aMC$@$NLO~\cite{Alwall:2014hca}.
The cross-sections are calculated by MadGraph5 at tree-level to estimate the constraints.  

We first consider the SM Higgs searches from VBF production by ATLAS at the 8 TeV \cite{Aaboud:2016cns} and 13 TeV  \cite{ATLAS:2016lgh} and also by CMS at the 8 TeV \cite{Rauco:2015elc} . The observed $95\%$ upper limit on SM Higgs cross section times the branching ratio is 4.1 pb from ATLAS and 4.6 pb from CMS at 8 TeV. For our benchmark point, the cross section for the process $pp \to j j \tilde K = 162$~pb with $p_{T,j} > 20$~GeV, $|\eta_j| < 5$. In order to obtain the rough idea about the constraint by comparing the LCH VBF search,  we simplify require $m_{jj}> 650$ GeV from the Madgraph parton-level simulation for the SM VBF Higgs process and for our $\tilde{K}jj$. This cut efficiency for $\tilde{K}jj$ is only 0.006 comparing to the cut efficiency on SM VBF process 0.23. Then the effective cross section after this cut for our benchmark point is only $\sigma(pp\to jj (\tilde K\to b\bar{b}))\sim 0.5$~pb by including the branching ratio of $\tilde K\to b\bar b$, which is smaller than the constraint from ATLAS \cite{Aaboud:2016cns} $0.94$pb and $1.06$pb from CMS \cite{Rauco:2015elc}. 

At 13TeV, ATLAS collaboration has explored SM Higgs in VBF production with an associated high energy photon in \cite{ATLAS:2016lgh}. The observed $95\%$ confidence level upper limit on the production cross section times branching ratio for a Higgs mass of 125 GeV is 4.0 times the Standard Model expectation. We use Madgraph to produce SM $Hjj\gamma$ and our model $\tilde{K}jj \gamma$, with both $H$ and $\tilde{K}$ decaying to $\bar{b}b$. At the parton level, we estimate the cross-section based on the basic cuts $p_T^j > 40$GeV, $p_T^\gamma >30$GeV, and $m_{jj}>800$GeV. After cuts and multiplying the corresponding $\bar{b}b$ BR, we found SM $Hjj\gamma \to \bar{b}bjj\gamma$ and $\tilde{K}jj\gamma \to \bar{b}bjj\gamma$ have cross-sections of about $4.5$ fb and $4.0$ fb respectively. Therefore, we conclude that our benchmark is not excluded by the constraints coming from the SM Higgs search in the VBF channel with an associated high energy photon.
 
Besides the VBF search, LHC also searched for SM Higgs via $ZH$ and $W^\pm H$ associated production. The constrain on such scenario is $\sigma(ZH)\times BR(H\to b\bar{b}) = 0.57^{+0.26}_{-0.23}$pb from ATLAS \cite{Aaboud:2017xsd}. For our model, the corresponding process is $pp\to \tilde Z\tilde K$, the cross section is suppressed by $\sin^2\alpha \sim 10^{-3}$, which is much smaller than the SM cross section.   
 
Before closing the section, we make some more comments on the  $2.9\sigma$ excess in the di-jet resonance searches at 13TeV CMS~\cite{CMS:2017dhi}, which motivated us to set $m_{\tilde K} = 115$ GeV as the benchmark point. This search is dedicated to look for new vector resonance $Z^\prime$, which only coupled to the SM quarks with universal vector-like coupling, and the largest deviation from the SM background only hypothesis is around $m_{Z^\prime}$ = 115 GeV with local significance $\sim 2.9 \ \sigma$. Comparing the observed $95\%$ CL upper limit cross section $\sim 1.05 \times 10^{4}$ pb for the $Z^\prime $ with the expected one  $\sim 4.5 \times 10^{3}$ pb, we can see that roughly one needs $5\times 10^3$pb to fit the excess. The cross-section in our benchmark point at tree level is $3.1\times10^3$ pb, which is capable to explain this excess. The search requires high $p_T$ $Z^\prime$ that the dijet merged into a single jet. Given that in our model, $\tilde{K}$ decays to $\bar{b}b$ and $\tilde{c}c$ at equal rate, it is interesting to analyze
what could be the significance had  CMS performed heavy flavor tagging, something not done in Ref.~\cite{CMS:2017dhi}. At 13TeV LHC~\cite{CMS:2017cbv}, CMS collaboration has looked for  the high $p_T$ fat jet with b-tagging in the inclusive $H+j$ measurement. The tagging efficiency is $33\%$ for $H\to (b\bar{b})$ as a fat jet and $1\%$ for mis-tagging efficiency from light flavor quarks. If applying b-tagging in $Z'$ resonance search in Ref.~\cite{CMS:2017dhi}, the increase in $S/\sqrt{B}$ is $50\% \times  33\% / \sqrt{1\%}\sim 1.6$ which is a moderate increase if the background error is statistic dominant. 

The CMS collaboration further used this high $p_T$ fat jet with b-tagging technique in related searches for the inclusive $H+j$ process with $H\to \bar{ b}b$, by requiring $p_T^H > 450 \GeV$~\cite{CMS:2017cbv}. 
The theoretical cross-section for $H(\bar{b}b) $ with $p_T^j >450$GeV is $31.7 \pm 9.5$ fb with $30\%$ uncertainty, while the measured value is $74 \pm 50$ fb. The mean value is therefore about 2.5~times the Higgs one, with an observed significance of $1.5\sigma$. No other significant resonances have been found. In our benchmark model, the cross-sections  after cut for $\tilde{K}j \to (bb)j$ and $\tilde{K}b\to (bb)b$ are about $41$ fb and $25$ fb respectively. Note that $Kj$ has a similar cross-section as $Hj$, and $m_{\tilde{K}} =115$ GeV in our benchmark. Moreover, with an extra b quark in $\tilde{K}b$, the mis-reconstruct, mis-combination and mis-tagging might result in a smaller contribution, thus we estimate its contribution should be less significant. The $m_{bb}$ distribution  in Fig. 4 of \cite{CMS:2017cbv} presents a broad excess that range from 105 GeV to 140 GeV
and therefore, although a dedicated experimental analysis must be performed, we conclude that the $\tilde{K}$ signal is compatible with the current experimental observations in this channel.  Higher luminosity LHC measurements in this channel are likely to provide the most effective way of probing this scenario.

\section{Dark Matter Search}
\label{sec:DM-search}

In this section, we will explore in detail the possibility of the neutral vector-like fermion $\chi_2$ being a  dark matter candidate. The interaction Lagrangian for $\chi_2$ in the mass basis at leading order of $\sin\alpha$ and $\cos\beta$ is 
\begin{align}
\mathcal{L}_{\chi_2}&\simeq -g_D \cos\alpha \tilde {K}_\mu \bar{\chi}_{2,R}\gamma^\mu \chi_{2,R} + g_D \sin\alpha \tilde{Z}_\mu \bar{\chi}_{2,R}\gamma^\mu \chi_{2,R}  \nonumber \\ 
&+\frac{m_{\chi_2}}{v_D}\bar{\chi}_2\chi_2\left(\frac{v}{v_D}\cos\beta H^0_1 + U_{23}H^0_2 -U_{33}H^0_3\right) + i \frac{m_{\chi_2} v}{v_D^2} \cos\beta A^0 \bar{\chi}_2\gamma_5 \chi_2 ,
\label{eq:DMint}
\end{align}
where $m_{\chi_2} = y_{\chi_2} v_D/ \sqrt{2}$. The  Majorana mass  term $\frac12 M_m\bar{\chi}_{2,L} \chi_{2,L}^c$ in \cref{eq:YukawaChis} will split the Dirac fermion into two Majorana fermions, which is similar to the inelastic DM setup considered in Ref.~\cite{TuckerSmith:2001hy}. In the Weyl fermion basis $\left(\chi_{2,L}, \chi_{2, R}^c \right)^T$,  the mass matrix  is given by: 
\begin{align}
M_\chi = \left(\begin{array}{cc}
 M_m & m_{\chi_2} \\
m_{\chi_2}  & 0 \end{array} \right) ,
\end{align}
where we assume $M_m , m_{\chi_2} >0$ without loss of generality.  This symmetric mass matrix can be diagonalized by an orthogonal rotation: 
\begin{align}
\left(\begin{array}{c} \chi_{2,L} \\ \chi^c_{2, R} \end{array}\right)
= U_{\chi_2} \left(\begin{array}{c} \eta_{1} \\ -i\eta_{2} \end{array}\right)
=\left(\begin{array}{cc}
c_{\chi_2} & s_{\chi_2} \\
-s_{\chi_2}  & c_{\chi_2} \end{array}
\right)  \left(\begin{array}{c} \eta_{1} \\ -i\eta_{2} \end{array}\right) ,
\end{align}
where $\eta_{1,2}$ are the mass eigenstates of two Majorana fermions and the factor $-i$ is to ensure the  Majorana masses of $\eta_{1,2}$ have the same value $m_{\chi_2}$ in the limit of $M_m = 0$.

In the  small Majorana mass limit $M_m \ll m_{\chi_2}$, the eigenstate masses are 
\begin{align}
m_{\eta_1} &= m_{\chi_2} +  \frac{M_m}{2}, \\
m_{\eta_2}& =   m_{\chi_2} -\frac{M_m}{2} ,
\end{align}
 where the mixing angle is given by:
 \begin{align}
  c_{\chi_2} &= \frac{1}{\sqrt{2}} + \frac{M_m}{4 \sqrt{2} m_{\chi_2}} \simeq \frac{1}{\sqrt{2}} ,\\
   s_{\chi_2}& = - \frac{1}{\sqrt{2}} + \frac{M_m}{4 \sqrt{2} m_{\chi_2}} \simeq  -\frac{1}{\sqrt{2}}.
 \end{align} 
 For large Majorana mass $M_m \gg m_{\chi_2}$, the eigenstate masses are 
 \begin{align}
 m_{\eta_1}&= M_m + \frac{m_{\chi_2}^2}{M_m}, \\
 m_{\eta_2} &= \frac{m_{\chi_2}^2}{M_m},
 \end{align}
which is a typical see-saw mass, with the mixing angle $s_{\chi_2} = - m_{\chi_2}/(M_m) \ll 1$. With the mixing angle we can rewrite the light Majorana DM $\eta_2$ back into its Dirac form,
 \begin{align}
 \chi'_2=\left(\begin{array}{c} i\eta_2 \\  -i\eta_2^\dag \end{array}\right),
 \end{align}
and also the interaction Lagrangian as follows: 
\begin{align}
\mathcal{L}_{\chi'_2}\simeq & -g_D \cos\alpha ~c^2_{\chi_2} \tilde {K}_\mu \bar{\chi'}_{2}\gamma^\mu \gamma_5\chi'_{2} + g_D \sin\alpha ~c^2_{\chi_2} \tilde{Z}_\mu \bar{\chi'}_{2}\gamma^\mu \gamma_5\chi'_{2}  \\ \nonumber
&+\frac{m_{\chi_2}}{v_D} 2s_{\chi_2}c_{\chi_2}\bar{\chi'}_2\chi'_2\left(\frac{v}{v_D}\cos\beta H^0_1 + U_{23}H^0_2 -U_{33}H^0_3\right) + i2s_{\chi_2}c_{\chi_2} \frac{m_{\chi_2} v}{v_D^2} \cos\beta A^0 \bar{\chi'}_2\gamma_5 \chi'_2 .
\label{eq:DMint2}
\end{align}
We can simplify it by
\begin{align}
\mathcal{L}_{\chi'_2} \simeq 
\left\{
\begin{array}{ll}
 \frac{-g_D }{2}\bar{\chi'}_{2}\gamma^\mu \gamma_5\chi'_{2} \left( \cos\alpha\tilde {K}_\mu + \sin\alpha \tilde{Z}_\mu \right) - \frac{m_{\chi_2}}{v_D} \bar{\chi'}_2\chi'_2U_{33}H^0_3   & (M_m \ll m_{\chi_2}) \\
-g_D\bar{\chi'}_{2}\gamma^\mu \gamma_5\chi'_{2}\left( \cos\alpha\tilde {K}_\mu + \sin\alpha \tilde{Z}_\mu \right)  &(M_m \gg m_{\chi_2})
\end{array}\right.,
\end{align}
where we keep only the leading order interactions in $\mathcal{O}(M_m/m_{\chi_2})$ or $\mathcal{O}(m_{\chi_2}/M_m)$. In this  following subsections, we will discuss the phenomenology of Dirac and Majorana DM separately.

\subsection{Dirac Dark Matter} 
We first consider the case of pure Dirac dark matter, whose Lagrangian is listed in Eq.~(\ref{eq:DMint}). We will study the condition to obtain the correct relic abundance and explore the dark matter limits from indirect detection, direct detection and collider searches.

\subsubsection{DM annihilation} 
We first calculate the  $\chi_2 \chi_2$ annihilation cross sections. The DM annihilation $\bar{\chi}_2 \chi_2 \to \bar{f}f$ is an s-channel process, mediated by $\tilde{K}$, $\tilde{Z}$, $H_{1,2,3}^0$ and $A^0$. From Eq.~(\ref{eq:DMint}), only processes with $b\bar{b}$($c\bar c$) final states and mediated by $\tilde{K}$ and $H_3^0$ are not suppressed by small mixing angle $\sin\alpha$ and $\cos\beta$. Given that the Yukawa couplings between $H_3^0$ and $b,c$ quarks are much smaller than 1, we conclude that the dominant DM annihilation process is $\bar{\chi_2} \chi_2\to \tilde{K}^{*} \to \bar{b}b,\bar{c}c $ with annihilation cross-section
\begin{align}
(\sigma v)_{\chi_2\bar{\chi}_2 \to \bar{q} q}^{q=b,c} & = \frac{g_D^4\sqrt{1-\frac{4 m_q^2}{s}} }{8 \pi 
\left( (s-m_{\tilde{K}}^2)^2  +  m_{\tilde{K}}^2 \Gamma_{\tilde{K}}^2\right)  } 
\left[ s-m_q^2 + m_{\chi_2}^2 
\left(-1 + \frac{m_q^2 (4 m_{\tilde{K}}^4 - 6 m_{\tilde{K}}^2 s + 3 s^2)}{m_{\tilde{K}}^4 s} \right) \right]
\nonumber \\
& \approx  \frac{g_D^4}{8 \pi \left( (s-m_{\tilde{K}}^2)^2  +  m_{\tilde{K}}^2 \Gamma_{\tilde{K}}^2\right)} (s - m_{\chi_2}^2) 
\end{align}
where we have neglected the quark mass in the second line. For the annihilation at freeze-out, it needs to be averaged over thermal distribution of DM, while for annihilation today, it only needs the substitution $s =  4 m_{\chi_2}^2$.

\begin{figure}
  \includegraphics[width=0.46 \columnwidth]{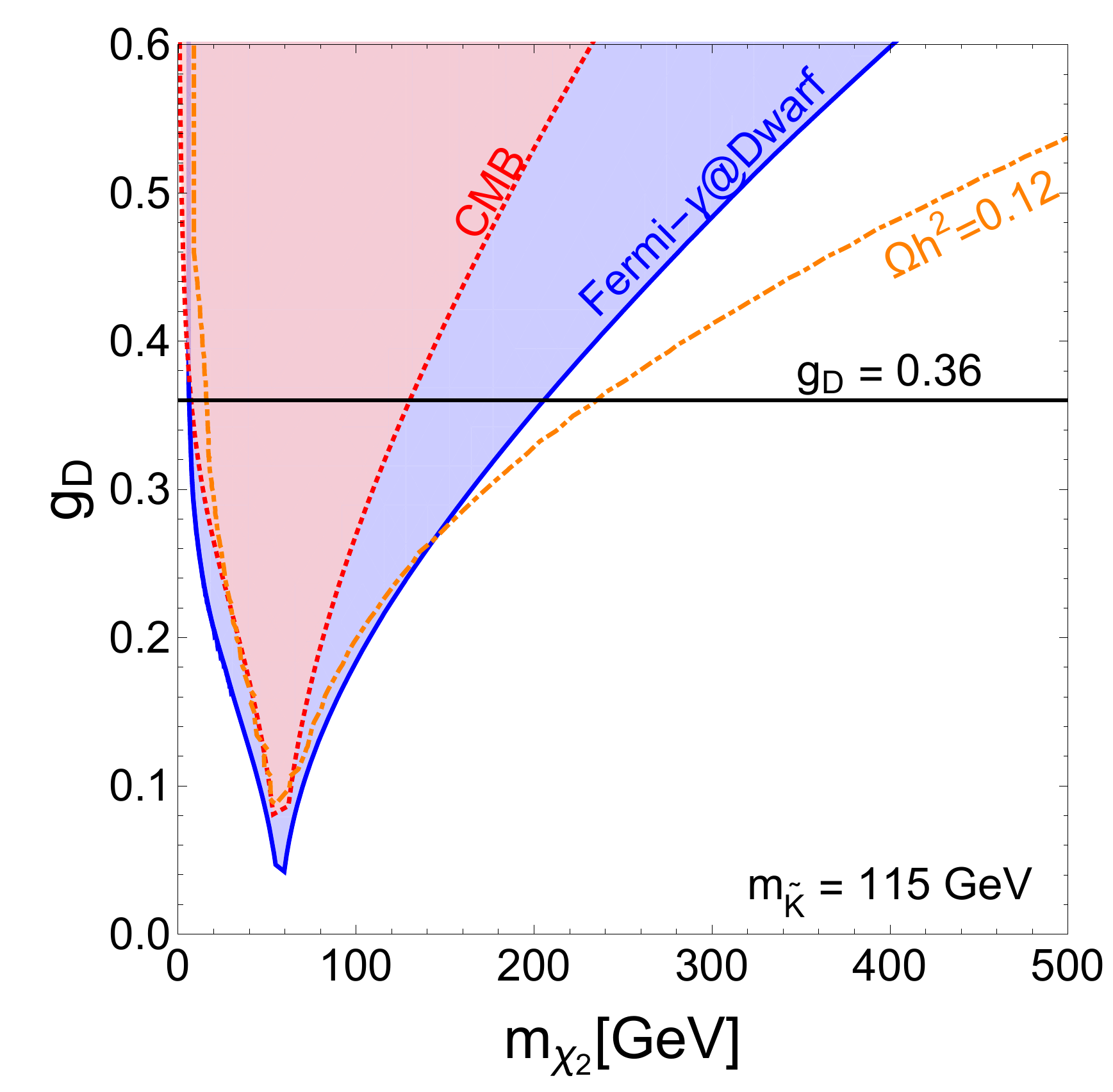} 
  \includegraphics[width=0.46 \columnwidth]{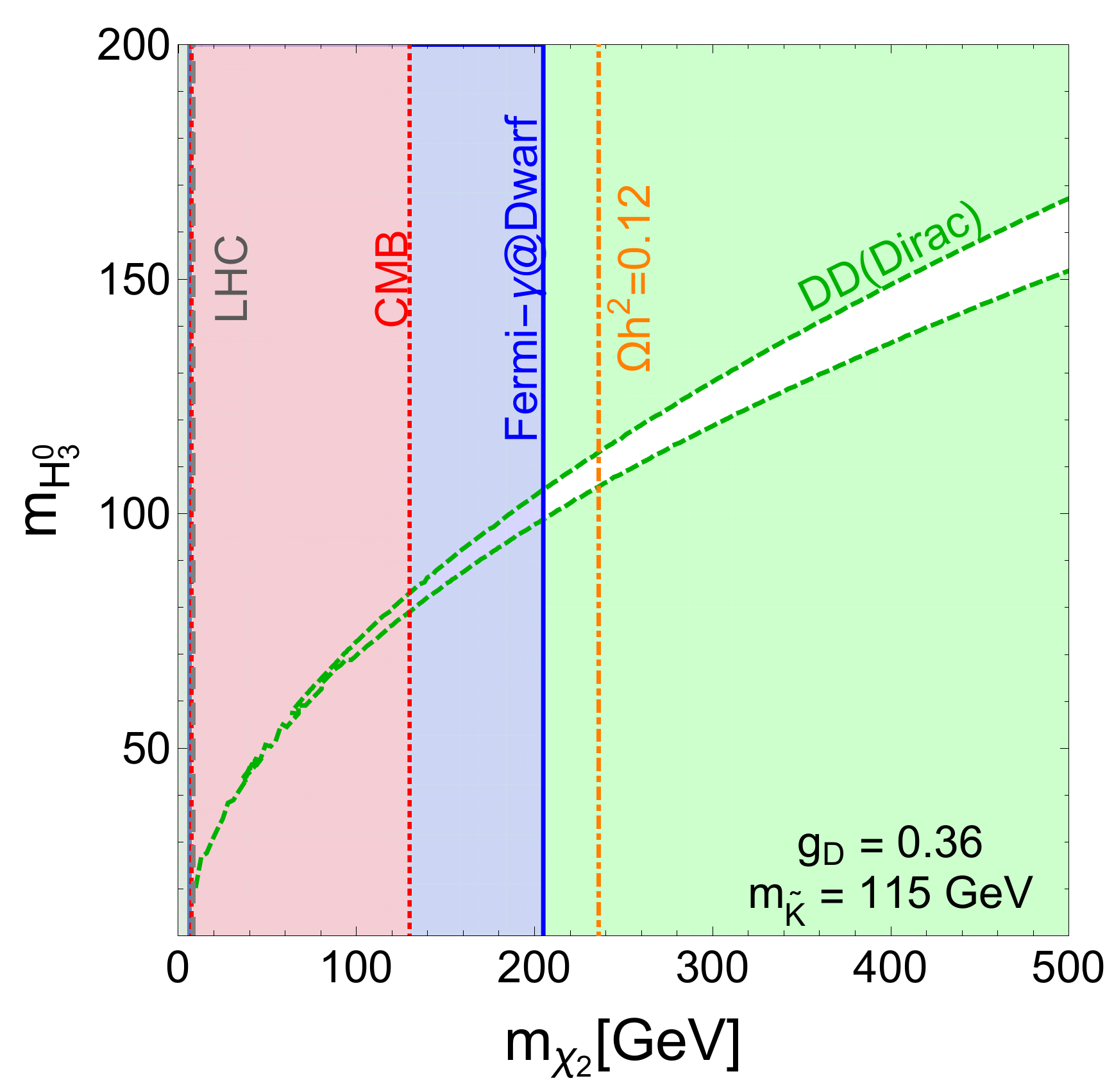}
  \caption{ The constraints on Dirac dark matter parameter space in $m_{\chi_2}$-$g_D$ plane (\textit{Left}) and $m_{\chi_2}$-$m_{H_3^0}$ plane (\textit{Right}). The orange line correspond to parameters that lead to the right relic abundance $\Omega h^2 =0.12$. The red shaded region gives the CMB limits \cite{Ade:2015xua}, while the blue shaded region gives the most stringent gamma-ray limits from Fermi observation in dwarf galaxies \cite{Ackermann:2015zua, Fermi-LAT:2016uux} (labeled as ``Fermi $\gamma$ at Dwarf galaxies"). The green area is excluded by Xenon1T \cite{Aprile:2017iyp} for benchmark point parameters. The gray line is limits from jets+MET with 1 b-jet tagging at 13TeV CMS \cite{Sirunyan:2017kiw}.
   }
  \label{fig:relic}
\end{figure}

To reproduce the right relic abundance $\Omega h^2 = 0.12$ \cite{Ade:2015xua}, the thermal averaged cross-section for Dirac fermion DM is about $6\times 10^{-26} \text{cm}^3/\text{s}$. In Fig.~\ref{fig:relic}, we plot the contours (the orange line) in the $m_{\chi_2} - g_D$ plane, which gives the  right relic abundance  for our benchmark scenario $m_{\tilde K} = 115$ GeV.  If we further choose $g_D= 0.36$ as our benchmark point, we obtain two solutions for the DM mass, $m_{\text{DM}}$ =14 GeV or 236 GeV, which can satisfy the relic abundance requirement.

\subsubsection{ DM indirect detection} 

The Dirac fermion DM $\chi_2$ annihilation to $\bar{b}b$ and $\bar{c}c$ have equal rate, with total annihilation cross-section leading to right relic abundance for DM mass $14~(236)$ GeV. Since the annihilation is s-wave, the final state particles from DM annihilation will inject energy into primordial plasma which would delay recombination and thus leave observable imprints in the Cosmic Microwave Background (CMB)~\cite{Adams:1998nr, Padmanabhan:2005es, Galli:2009zc, Slatyer:2009yq}. Given that energy injection efficiency of $\bar{b}b$ and $\bar{c}c$ are similar \cite{Slatyer:2015jla}, the constraint from CMB \cite{Ade:2015xua} is
\begin{align}
p_{\rm{ann}}(z)\equiv f(z)\frac{\left<\sigma v\right>}{m_{\chi_2}}<3.5\times 10^{-28} \rm{cm^3 s^{-1} GeV^{-1}},
\end{align}
Making use of the $f(z)$ function from \cite{Slatyer:2009yq}, we plot the excluded region (in red) in Fig.~\ref{fig:relic}, where we can see that the low mass benchmark $m_{\chi_2} = 14$~GeV is excluded, while  the high mass $m_{\chi_2} = 236$ GeV is still allowed .  

In addition, the Fermi-LAT gamma-ray observations of dwarf galaxies provide a constraint on the DM annihilation cross-sections based on final states \cite{Ackermann:2015zua, Fermi-LAT:2016uux}. For $\bar{b}b$ final states, this tells us that the DM mass should be larger than 100 GeV, i.e. $m_{\chi_2}\gtrsim 100$ GeV, in order to have the right thermal relic density. Since the photon spectrum from final state $\bar{b}b$ and $\bar{c}c$ are quite similar \cite{Ciafaloni:2010ti}, it again excludes the light DM benchmark but not for the heavy one. The gamma-ray observation from Galactic Center (GC) by Fermi-LAT gives constraint $m_{\text{DM}}\gtrsim 50$ GeV for $\bar{b}b$ final states~\cite{TheFermi-LAT:2017vmf}, which is less stringent than dwarf galaxies. There is also a gamma-ray constraint from the Virgo cluster \cite{Ackermann:2015fdi}, but is much weaker than the above two constraints. Therefore, in Fig.~\ref{fig:relic}, we only show the most stringent limits from Fermi dwarf galaxies observation in blue shaded area.

\subsubsection{ DM direct detection} 
In this section, we will consider the direct detection (DD) of $\chi_2$, which are related to the scattering between $\chi_2$ and nucleon. The sum of different flavor quark contribution inside nucleon from scalar mediator should be performed at the amplitude level and the results read:
\begin{align}
a_N = \left(\sum_{q=u,d,s} f_{\text{Tq}}^{(N)} \frac{a_q}{m_q} + \frac{2}{27} f_{\text{TG}}^{(N)} \sum_{q=c,b,t} \frac{a_q}{m_q}  \right)m_N ,
\end{align}
where $f_{\text{TG}}^{(N)}, f_{Tq}^{(N)}$ are the form factors  and $N=p,n$ is proton and neutron respectively. The quark form factors for proton are$f_{\text{Tu}}^{(p)} = 0.017 \pm 0.008$, $f_{\text{Td}}^{(p)} = 0.028 \pm 0.014$, $f_{\text{Ts}}^{(p)} = 0.040 \pm 0.020$, $f_{\text{TG}}^{(p)} \approx 0.91$~\cite{Junnarkar:2013ac, Hill:2011be} and for neutron are $f_{\text{Tu}}^{(n)} = 0.011$ $f_{\text{Td}}^{(n)} = 0.0273$, $f_{\text{Ts}}^{(n)} = 0.0447$, $f_{\text{TG}}^{(n)} \approx 0.917$~\cite{Belanger:2006is} (see also results from \cite{Alarcon:2011zs, Alarcon:2012nr}). In our model, the scattering between nucleon and $\chi_2$ are mediated by CP even scalars $H^0_{1,2,3}$, CP odd scalar $A^0$ and neutral gauge boson $\tilde{K}$ and $\tilde{Z}$. We will consider the scalar and vector contribution separately in the next two paragraphs.

\begin{figure}
  \includegraphics[width=0.23 \columnwidth]{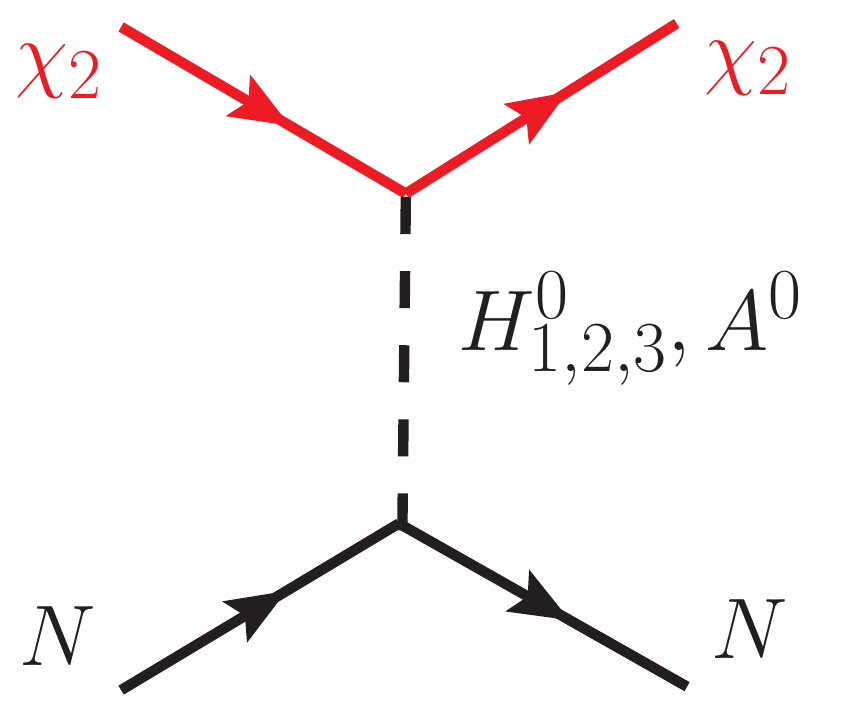} 
  \includegraphics[width=0.5 \columnwidth]{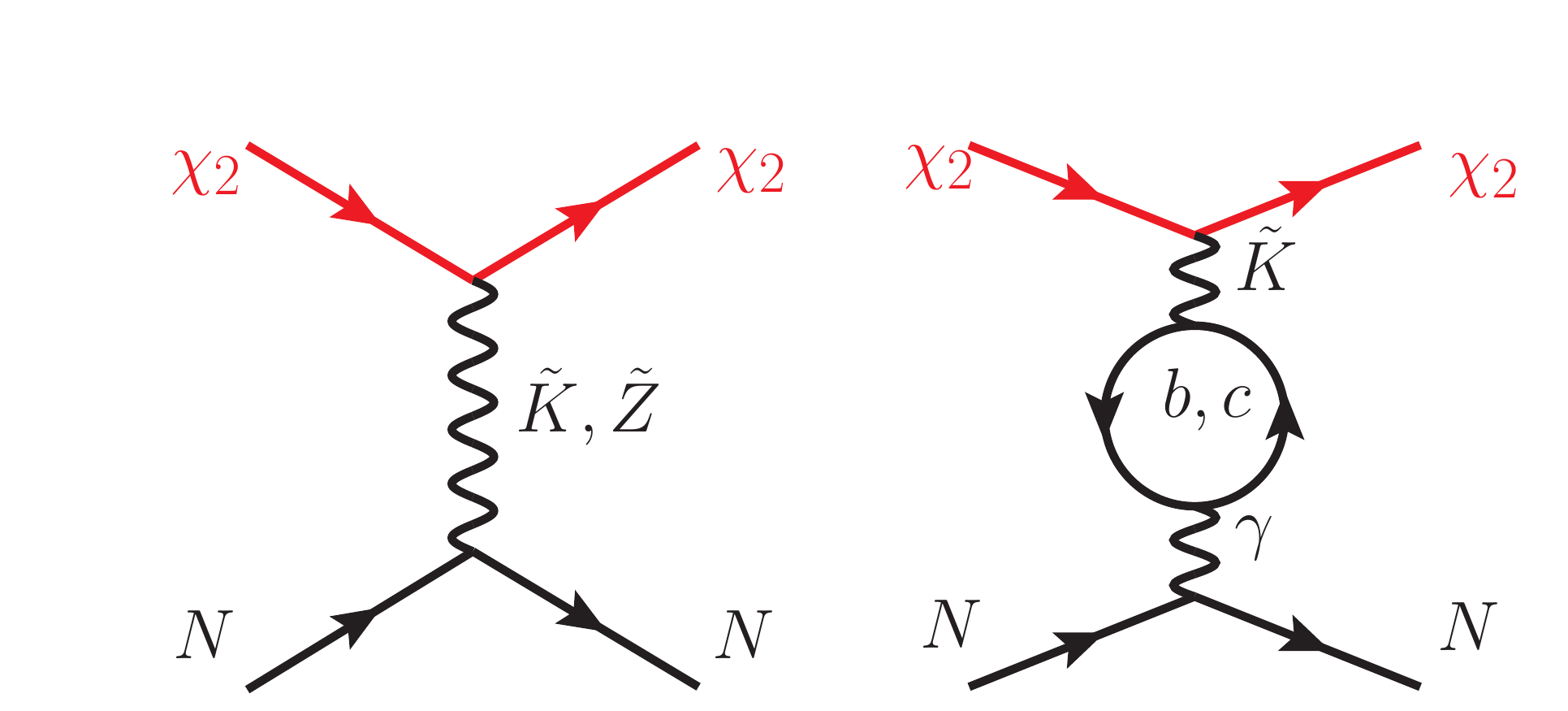}\\
  (a) ~~~~~~~~~~~~~~~~~~~~~~~~~~~~~~~~(b) ~~~~~~~~~~~~~~~~~~~~~~~~(c)
  \caption{The Feynman diagrams for Dirac $\chi_2$ scattering with nucleon. The figure (a) and (b) are mediated by scalars and vectors, while (c) is by kinetic mixing of gauge boson at 1-loop. The dominant contributions come from $H_3^0$ in (a) and $\tilde{K}, \tilde{Z}$ in (b).
   }
  \label{fig:feynman-DD}
\end{figure}

For scalar contribution shown in panel (a) of Fig.~\ref{fig:feynman-DD}, the scattering between $\chi_2$ and light flavor quark mediated via $H_2^0, H_3^0$ is suppressed by $\cot \beta$, where the suppression for $H_2^0$ is from $\chi_2$ Yukawa vertex and for $H_3^0$ is from light flavor $q$ Yukawa vertex, both from $U_{23}$ mixing. The $\chi_2$ and light quark scattering mediated by $H_1^0, A^0$ are suppressed by $\cot^2 \beta$. Therefore, only the scattering process with heavy flavor quark $b, c $ mediated by $H_3^0$, are not suppressed by scalar mixing $\sim \cot \beta$ and quark mixing angles. The leading contribution to spin-independent (SI) search is thus $H_3^0$ mediation via quark form factor $f_{\text{TG}}^N$, with the  amplitude $a_q$  proportional to $m_{\tilde b (c)}$. Given that $f_{\text{TG}}^{(p)} \approx f_{\text{TG}}^{(n)}$, the scalar contribution of $H_3^0$ are isospin universal.

For vector mediation by $\tilde{K}$ and $\tilde{Z}$, we only consider the vector-vector (V-V) fermion bilinear coupling contribution for SI interaction, shown in panel (b) and (c) in Fig.~\ref{fig:feynman-DD}. The reason is that vector-axial (V-A) and axial-vector (A-V) fermion bilinear are further suppressed by powers of velocity or momentum transfer, while axial-axial (A-A) interaction contributes to spin-dependent interaction which is less constraining than SI interaction~\cite{Bishara:2016hek, Escudero:2016gzx}. There are two kinds of contribution to the $\chi_2$ and nucleon V-V scattering. The first one is DM current couples to $J_Z$ current in SM,  $J_{\chi_2, \mu} J_{Z}^\mu$, which is suppressed by $\tilde{K}$, $\tilde{Z}$ mixing, shown as panel (b) in Fig.~\ref{fig:feynman-DD}. The second one is DM current interaction with $b,c$ quark currents mediated by kinetic mixing $\tilde{K}_{\mu\nu}B^{\mu\nu}$ where $B$ is hypercharge field, which is mixing suppression free, shown as panel (c) in Fig.~\ref{fig:feynman-DD}. Similarly like gluon form factor in nucleon for scalar interaction, the second one can induce a coupling to light quark in the nucleon via ElectroMagnetic current $\epsilon e J_{\text{EM}}$, by 1-loop contribution from $b,c$ quark. The kinetic mixing parameter $\epsilon \approx g_D g'/\left(16 \pi^2 \right) \sim 10^{-3}$, while given $\sin\alpha \sim 0.03$ in our benchmark, we have $ \sin\alpha g/c_w \gg \epsilon e$. Thus, it is reasonable to ignore the contribution shown in Fig.~\ref{fig:feynman-DD} (c).     

In the non-relativistic and heavy DM limit, both scalar mediation and V-V mediation have the fermion bilinear $\bar{\chi_2} (1+\gamma^0) \chi_2 \bar{N}N/2$, and we can calculate the SI cross-section for $\chi_2$ scattering with nucleon $N$~\cite{Bishara:2016hek}. Note that the $J_Z$ current involves an isospin violating coupling that $f_p  = \frac{g}{4 c_w} (1- 4 s_w^2)$ and $f_n  = \frac{g}{4 c_w} (- 1 )$, therefore we should average over proton and nucleon in the nuclei. The averaged SI cross-section for $\chi_2$ and nucleon is
\begin{align}
\sigma_N^{\text{SI}} = \frac{\mu_N^2}{2916 \pi v_D^4} 
\left(8 f_{\text{TG}}^{(N)} \frac{m_{\chi_2} m_N}{m_{H_3^0}^2} - 27 g_D \sin\alpha \frac{v_D^2 (m_{\tilde{K}}^2 - m_{\tilde{Z}}^2 )}{m_{\tilde{K}}^2 m_{\tilde{Z}}^2} \left(f_n \left(1-\frac{Z}{A} \right)  + f_p \frac{Z}{A} \right) \right)^2  ,
\label{eq:SI-1}
\end{align}
where $\mu_N^2 = m_{\chi_2} m_N/(m_{\chi_2}+m_N)$ is the DM-nucleon reduced mass. 
From Eq.~(\ref{eq:SI-1}), we see that $\tilde{K}$ and $\tilde{Z}$ contribution cancels each other due to mass mixing effect. In our benchmark point $g_D=0.36$, $\sin\alpha=-0.03$,  and $m_{\tilde K}=115 $GeV, with $|f_p| \ll |f_n| $ and $f_n < 0$, we found that the scalar mediated amplitude and vector mediated amplitude interfere destructively.  If $m_{H_3^0} \sim v_D$, then the vector contribution dominates, and $\sigma_N^{\text{SI}}$ does not depend on $m_{H_3^0}$ and $m_{\chi_2}$, and is around $5 \times 10^{-44} \text{cm}^2$ for our benchmark scattering with Xenon. 
Note that this contribution would be even larger, but thanks to the cancellation between $\tilde{K}$ and $\tilde{Z}$, the vector contribution gets a reduction of about $1/25$ already. The current limits on $\sigma_N^{\text{SI}}$ are from PANDAX-II, LUX and Xenon1T~\cite{Cui:2017nnn, Akerib:2016vxi, Aprile:2017iyp}, and for DM mass around $10 \sim 100$ GeV is of the order of a few $10^{-46} \text{cm}^2$.  Therefore, in order to satisfy the DD bounds, a cancellation between the vector and scalar contributions is required. We show the constraint from Xenon1T in the right panel of Fig.~\ref{fig:relic}. The green area is excluded by Xenon1T with our benchmark point.   For the values allowed by indirect detection, $m_{\chi_2} =  236$ GeV, the allowed region for $m_{H^0_3}$ is $103-116$ GeV. Therefore, if $\chi_2$ is a Dirac fermion, we need to tune the mass of $H_3^0$ to avoid direct detection limit with a level of $\sim 10\%$ tuning in mass. 

\subsubsection{ DM searches at the LHC} 
We start by analyzing DM searches at the LHC with mono-jet process $pp \to j \tilde{K} \to j (\chi_2\bar{\chi}_2)$. The cross section is $\sigma(j\tilde {K}) \times {\rm{BR}}(\tilde K \to \chi_2\bar{\chi}_2)$ for on-shell $\tilde{K}$ production if $m_{\chi_2} < m_{\tilde K}/2$, or $\sigma(pp\to j \chi_2\bar{\chi}_2)$ which is suppressed by 3 body phase space. 
We first consider the constraint when $m_{\chi_2} < m_{\tilde K}/2$, which is, however, in tension with indirect detection limits. Then the branching ratio of $\tilde K \to \chi_2 \bar{ \chi}_2$ varies from 0.14 to 0 when the mass $m_{\chi_2}$ is varied from $m_{\chi_2} = 0$ to $m_{\tilde K}/2$. Taking a benchmark point $m_{\tilde K} = 115$ GeV, $g_D =0.36$, then the cross section of $p p\to j \tilde K$ at 13TeV LHC is $1.3\times 10^3$ pb, and at 8TeV LHC is 960 pb.  Then we consider  the jet plus MET constraints from LHC 13TeV data with integrated luminosity $36 \text{fb}^{-1}$~\cite{Aaboud:2017phn, Sirunyan:2017cwe, Sirunyan:2017kiw}. The ATLAS collaboration~\cite{Aaboud:2017phn} studied the mono-jet limits for vector and axial vector mediator between SM quarks and DM. Their inclusive region (IM1) requires $\slashed{E}_T > 250 $ GeV which gives $95\%$ C.L. constraints on cross-section smaller than $0.53$ pb. We calculated parton level process $j + \tilde{K}$ with a requirement $P_t^j > 250$ GeV, leading to cross-section of about $2.4$ pb at 13TeV. Then we obtain a  constraint on the branching ratio ${\rm {BR}}(\tilde K \to \chi_2\bar{\chi}_2) < 0.22 $, which is always satisfied in the low mass region of $m_{\chi_2}$. 
The limit is given in Fig.~\ref{fig:relic} as gray dashed line, showing that $m_{\chi_2} \lesssim 10$ GeV is excluded.  

Aside from  mono-jet process, the mono-X ($X = A/W/Z$) processes are also interesting to look for. However, in s-channel vector mediator type
models, usually the mono-jet channel provides the strongest limits~\cite{Brennan:2016xjh}. Multi-jets plus missing
energy processes have been considered in addition to mono-jet channel to constrain DM simplified models. The usual expectation is that the two type of constraints have 
comparable limits, which is the case for s-channel vector mediator type models, ~\cite{Liew:2016oon}. 
This is different from scalar and pseudo-scalar mediators
with couplings to quarks which are proportional to quark masses, for which multi-jets process provides stronger limits~\cite{Buchmueller:2015eea}. 
The reason is that the production of scalar mediators is typically dominated by gluon fusion, which leads to more events with
higher jet multiplicity~\cite{Haisch:2012kf, Fox:2012ru, Haisch:2013ata}.

For the other two CMS multi-jet plus MET searches \cite{Sirunyan:2017cwe, Sirunyan:2017kiw}, the constraints should provide similar limits as mono-jet searches~\cite{Liew:2016oon}. For the case with no b-tagging, we have checked the signal bin $1$ and $2$ in Table B.1 of  Ref.~\cite{Sirunyan:2017cwe}. However, we found the constraint is weaker than mono-jet search~\cite{Aaboud:2017phn}, probably because this is a parton level estimation. Adding parton shower and detector simulation should bring a conclusion close to Ref.~\cite{Liew:2016oon}.
Given the $K_\mu$ are not universally coupled to all quarks but couple specifically with b quark and c quark, it is natural to pay special attention to signal regions with b-jet tagging. The CMS sbottom search~\cite{Sirunyan:2017kiw} looks for di-jet plus MET with b-tagging. 
The most prominent production mode in our model is a single bottom quark in association with $\tilde{K}$ that correspond to what is called the ``Compressed" search region. We checked the two Bins with $\slashed{p}_T $ within $\left[250, 300 \right]$ and $\left[300, 500 \right]$ with 1 b-jet and $H_T^b <100$ GeV requirements. The $95\%$ C.L. limits on the cross-section are about $14$~fb and $18$~fb. We calculated the cross-section from parton level analysis respectively, and the corresponding cross sections after cut are $7$ fb and  $8$ fb for our benchmark $m_{\tilde K} = 115$ GeV, respectively. Therefore it does not provide an efficient constraint on ${\rm {BR} }(\tilde {K} \to \chi_2\bar{\chi}_2) $. For the CMS multi-jet plus MET search~\cite{Sirunyan:2017cwe}, we have checked the b-tagging signal bins 11, 12, 21 and 22 which has $N_{\text{b-jet}} = 1, 2$, and found the sbottom search induced constraints~\cite{Sirunyan:2017kiw} improve but are still weaker than the ones coming from mono-jet searches~\cite{Aaboud:2017phn}.

Then we consider the case with $m_{\chi_2} > m_{\tilde K}/2$. The largest cross-section that may be obtained when $m_{\tilde K}=115$ GeV is for $m_{\chi_2} \simeq 58$~GeV. We get off-shell $\tilde{K}$ produced $j\chi_2\bar{\chi}_2 $ cross-section to be $\sim 0.01$ pb for $g_D=0.36$ and $\sin\alpha=-0.03$, where we only cut on $p_T^j > 250$ GeV. This is safe from the constraints at 13TeV LHC~\cite{Aaboud:2017phn} that cross-section after all cuts should be smaller than 0.57 pb. When $m_{\chi_2}$ is larger, the limits are even weaker due to smaller cross-section. We also check the process $jj \chi'_2\bar{\chi'}_2$ with our benchmark setup and found it is  even safer from Ref.~\cite{Sirunyan:2017kiw}.
For $m_{\chi_2} > m_{\tilde K}/2$, it is in general safe from the limits, due to small heavy quark PDF, 3-body phase space and off-shell suppression.

\subsection{Majorana Dark Matter}

If the mass of dark matter has contribution from a Majorana mass, the interactions between $\chi'_2$ and other particles are listed in Eq.~(\ref{eq:DMint2}). In this section, we will discuss the phenomenology of such Majorana DM, including constraints from dark matter relic abundance, direct detection, indirect detection and collider searches.

\subsubsection{DM annihilation}
The dominant annihilation process of $\chi'_2\bar{\chi'}_2 \to f\bar{f}$ is also mediated by $\tilde K$, after considering the mixing angle and Yukawa coupling suppression. The annihilation cross section is 
\begin{eqnarray}
(\sigma v)_{\chi'_2\bar{\chi'}_2 \to \bar{q} q}^{q=b,c} & =&\frac{g_D^4 c^2_{\chi_2}\sqrt{1-\frac{4m_q^2}{s}}}{4\pi \left[(s-m^2_{\tilde K})^2 + m^2_{\tilde K} \Gamma^2_{\tilde K} \right]} \left\{ \left(s-4m^2_{\tilde \chi_2}\right) + m^2_{q} \left[2m^2_{\chi'_2}\left( \frac{5m^4_{\tilde K} - 6m^2_{\tilde K } s+3s^2}{sm^4_{\tilde K}}\right) -1\right]\right\} \nonumber 
\label{eq:Ann-Mj}
\end{eqnarray}
From the annihilation formula in Eq.~(\ref{eq:Ann-Mj}), we can find out there are two contributions, one is p-wave suppressed which is proportional to $(s-4 m_{\tilde{\chi}_2}^2) \propto m^2_{\chi'_2} \vec{v}^2$, the other is helicity suppressed proportional to the quark mass. If we want to consider the annihilation cross section at freeze out, when the temperature is around $T_f\sim m_{\chi'_2}/20$, the dominant contribution comes from $m^2_{\chi'_2} \vec{ v}^2$ term. After thermal averaging, we find out the necessary dark matter mass to obtain the observed relic abundance is about 22 GeV or 142 GeV, as shown in orange lines in Fig.~\ref{fig:Constraint-DM-MJ}. Note this result is for large mass splitting $M_m \gg m_{\chi_2}$. If $\Delta m \equiv m_{\eta_1} - m_{\eta_2}$ is within $\Delta m / m_{\eta_2} \lesssim T_f/m_{\eta_2} \sim 0.05 $, the co-annihilation with $\eta_1$ will give a result close to Dirac DM result.

\subsubsection{DM indirect detection}

From the annihilation cross-section listed in Eq.~(\ref{eq:Ann-Mj}), both contribution from p-wave and quark mass terms are very tiny for annihilation today. Therefore, there is no constraint from indirect detection. 

\subsubsection{DM direct detection}
\begin{figure}
  \includegraphics[width=0.5 \columnwidth]{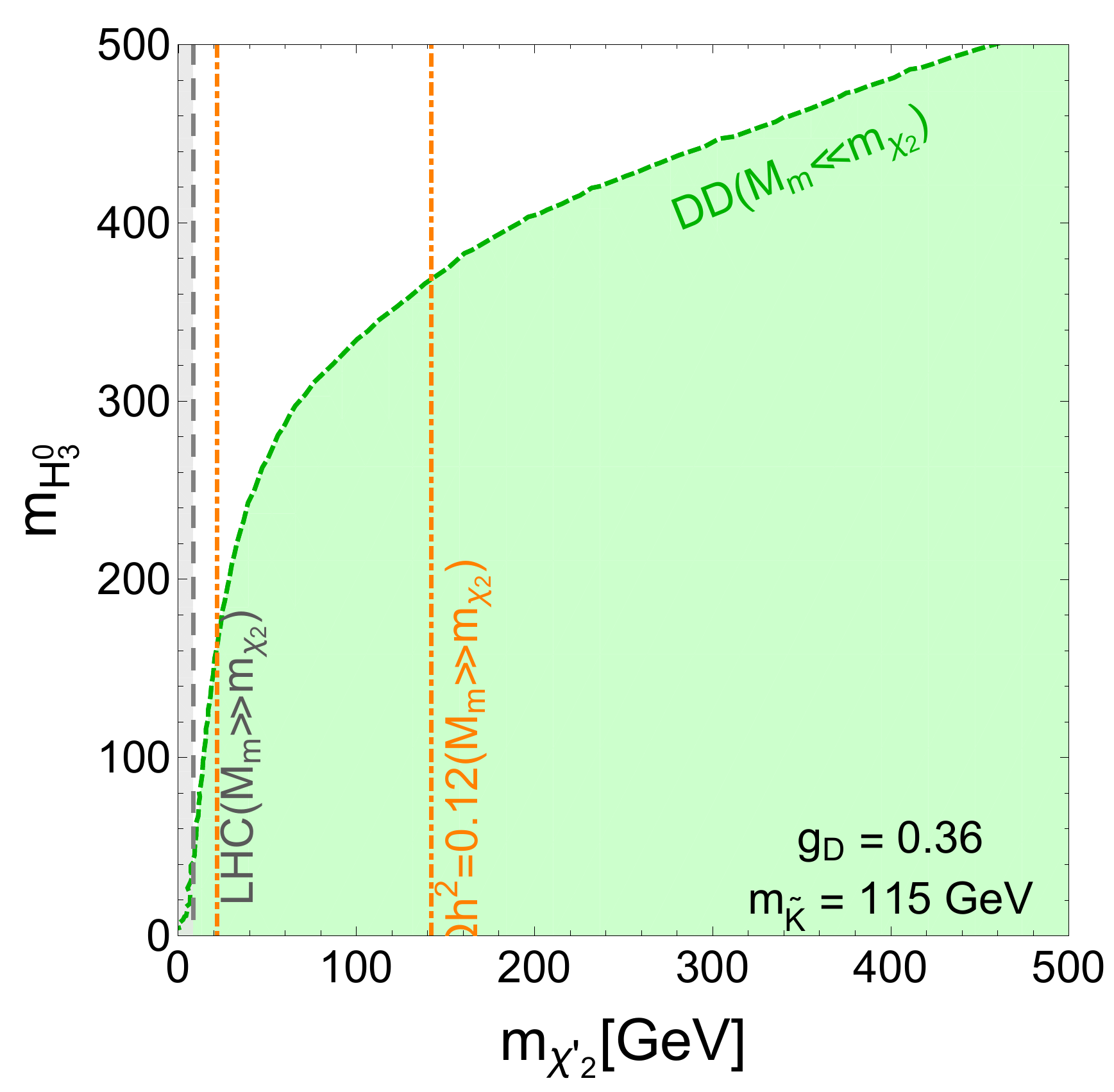} 
  \caption{The constraints for Majorana dark matter from LHC searches, direct detection and relic abundance. The labels are similar as in Fig.~\ref{fig:relic}. }
  \label{fig:Constraint-DM-MJ}
\end{figure}

 Comparing the Lagrangian between Dirac case, Eq.~(\ref{eq:DMint}), and Majorana case, Eq.~(\ref{eq:DMint2}), we find out the coupling of dark matter to gauge bosons $\tilde K$ and $\tilde Z$ are different. In the Majorana case, the vector coupling becomes 
 $\gamma_\mu \gamma_5$ which induces spin dependent (SD) interaction, or velocity (momentum transfer) suppressed SI interaction. Therefore, the cross-section of vector mediated processes is very small and can be ignored. There are vector coupling between DM $\eta_2$ and its excited state $\eta_1$, but it will be irrelevant if mass splitting is larger than $\mathcal{O}(100)$ keV. For the scalar part, if $M_m \gg m_{\chi_2}$, $s_{\chi_2} = - m_{\chi_2}/M_m$ is very tiny so we can ignore the cross-section. Therefore, there are no constraints from direct detection. In the small splitting case $M_m \ll m_{\chi_2}$, $s_{\chi_2}\sim c_{\chi_2} \sim 1/\sqrt{2}$, then the coupling between $\chi'_2$ and Higgs are similar as in the Dirac case, and hence the scattering cross-section between $\chi'_2$ and nucleon is
\begin{align}
\sigma_N^{\text{SI}} = \frac{16\mu_N^2}{729 \pi v_D^4} 
\left(f_{\text{TG}}^{(N)} \frac{m_{\chi'_2} m_N}{m_{H_3^0}^2} \right)^2  .
\label{eq:SI-2}
\end{align}
With the scattering cross-section, we give the constraints on $m_{H^0_3}$-$m_{\chi'_2}$ plane, which are shown as the green area in Fig.~\ref{fig:Constraint-DM-MJ}. If $m_{H^0_3}$ is large enough, the cross-section will be very tiny. 

\subsubsection{DM search at LHC}

As previously discussed, the search for dark matter at LHC is dominantly via the interaction between dark matter and $\tilde K$. The branching ratio of ${\rm{BR}}(\tilde K)\to \chi'_2\bar{\chi'}_2$ goes from $1/7$ to 0 when $m_{\chi'_2} < m_{\tilde K}/2$ for $M_m \gg m_{\chi_2}$. Since the cross-section of $j \tilde {K}$ and $jj \tilde{K}$ at 13~TeV LHC does not change with respect to the Dirac case, 
if we consider the case $m_{\chi'_2}< m_{\tilde K}/2$, then the constraints on invisible decay branching ratio is the same as the Dirac case. After combining the constraints from ATLAS and CMS, we can still make use of the limit ${\rm{BR}}(\tilde K)\to \chi'_2\bar{\chi'}_2 < 0.14$ leading to  a constraint on $m_{\chi'_2} \lesssim 8.5 $GeV for $M_m \gg m_{\chi_2}$. We show the LHC constraints as a gray area in Fig.~\ref{fig:Constraint-DM-MJ}, which does exclude low mass DM benchmark. 
For $M_m \ll m_{\chi_2}$, the branching ratio ${\rm{BR}}(\tilde K)\to \eta_{1,2} \eta_{1,2}$ are similar as ${\rm{BR}}(\tilde K)\to \chi'_2\bar{\chi'}_2$ for $M_m \gg m_{\chi_2}$, where each channel contributes approximately $1/4$. But for sizable mass splitting between $\eta_1$ and $\eta_2$, the limits will be weaker because some channels $\eta_{1} \eta_{2}$ or $\eta_{1} \eta_{1}$ may not be kinematically accessible.

If $m_{\chi'_2}> m_{\tilde K}/2$, for the search of jet+MET at ATLAS, we compare our cross-section $\sim$ 0.01 pb after cut $p_T^j>250$ GeV to the constraint at 13 TeV LHC which is 0.57 pb \cite{Aaboud:2017phn}. It shows that off-shell $\tilde{K}$ is very safe from limits from mono-jet searches. We also check the process $jj \chi'_2\bar{\chi'}_2$ with our benchmark setup and found it is  even safer from the constraints obtained in Ref.~\cite{Sirunyan:2017kiw}.

\section{Conclusions}
\label{sec:Conclusions}

In this article we have studied a gauge extension of the SM that allows to explained the observed deviation of the forward backward
asymmetry of the bottom-quark with respect to the expected value in SM. The new gauge boson should be neutral and
should couple to only right-handed bottom and charm quarks at tree-level. Coupling to the other fermions are only induced via 
mixing, which should be small and fixed by the relation $g_D \sin\alpha \simeq -0.011$ in order to obtain the right modification of
the right-handed bottom quark $Z$ coupling. Additional Higgs states are necessary for the realization of this scenario, but their
signatures are too weak to ensure detection at the LHC. Moreover,  we showed that, provided the new gauge boson mass is 
about $\sim 115$~GeV this model is consistent with all experimental constraints. The new gauge boson decays mostly into bottom and
charm jets and could provide an explanation of an observed di-jet excess in boosted topologies at the CMS experiment. 

Cancellation of anomalies in this model leads to the presence of a charged, vector-like lepton singlet state, as well as a 
vector-like neutral state that serves as a good DM candidate.  If it is a pure Dirac fermion, we can obtain the right relic abundance when its mass is around $14$~or~236 GeV. The indirect detection for DM annihilation induced gamma-rays rules out the low mass DM benchmark, but keeps the high mass benchmark intact.  The direct detection excludes the heavy Dirac $\chi_2$ benchmarks, unless  a $10\%$ fine tuning in $H_3^0$ mass is applied. If $\chi_2$ is split into two Majorana fermions, as is naturally the case, the direct detection constraint is easily evaded for large enough $m_{H^0_3} > 400$~GeV. We also can  get the right relic abundance for a mass $m_{\chi'_2} \sim 22$ or 142 GeV. There are no indirect detection limits because the annihilation cross-section at low temperatures is highly suppressed. The LHC searches does not rule out the both DM benchmark points, but is marginal for low mass DM benchmark point.

\section*{Acknowledgments}
Work at University of Chicago is supported in part by U.S. Department of Energy grant number DE-FG02-13ER41958. Work at ANL is supported in part by the U.S. Department of Energy under Contract No. DE-AC02-06CH11357.  The work  of CW was partially performed at the Aspen Center for Physics, which is supported by National Science Foundation grant PHY-1607611.   We would like to thank Zhen Liu, LianTao Wang for useful discussions and  comments. 
JL acknowledges support by Oehme Fellowship.

\appendix
\section{CP-even Higgs masses and mixing matrix}
\label{sec:CPevenHiggs}

In this section, we list a more detailed expression for CP-even Higgs mass and mixing matrix. The mass eigenvalues for CP-even Higgs in small $\cot\beta \equiv v_1/v_2$ and $\lambda_6$ expansion are given below,
\begin{align}
m^2_{H^0_1} &= -\frac{\mu_8 v_D \tan\beta}{\sqrt{2} }
- \cot\beta \frac{\mu_8 (v_2^2+v_D^2)}{\sqrt{2}v_D} + 2 (\lambda_2+\lambda_3)
\cot^2\beta v_2^2
,   \\
m^2_{H^0_2} & = 2 \lambda_2 v_2^2 
+ \cot^2\beta v_2^2 
\left( \frac{  \mu_8^2}{ \lambda_2 v_2^2 - \lambda_3 v_D^2} -  2 \lambda_2 \right)
+ \lambda_6^2 \frac{  v_2^2 v_D^2}{ 2\lambda_2 v_2^2 - 2\lambda_3 v_D^2}
+\lambda_6 \cot\beta \frac{\sqrt{2} \mu_8 v_D v_2^2}{\lambda_2 v_2^2 - \lambda_3 v_D^2}
,   \nonumber  \\
m^2_{H^0_3} & = 2 \lambda_3 v_D^2 
+ \cot^2\beta v_2^2 
\left( \frac{ - \mu_8^2}{ \lambda_2 v_2^2 - \lambda_3 v_D^2} -  2 \lambda_3 \right)
- \lambda_6^2 \frac{  v_2^2 v_D^2}{ 2\lambda_2 v_2^2 - 2\lambda_3 v_D^2}
- \lambda_6 \cot\beta \frac{\sqrt{2} \mu_8 v_D v_2^2}{\lambda_2 v_2^2 - \lambda_3 v_D^2} . \nonumber
\label{eq:M2largemu8}
\end{align}

The mixing matrix in Eq.~(\ref{eq:Umix}) are given in the more detailed expressions below,
\begin{align}
U_{11} &= 1 -\cot^2\beta \frac{v_2^2+v_D^2}{2v_D^2}   \\
U_{12} &= \cot\beta + \mathcal{O}(\cot^2\beta) + \mathcal{O}(\cot\beta \lambda_6)  \nonumber \\
U_{13} &= \cot\beta \frac{v_2}{v_D}+ \mathcal{O}(\cot^2\beta) + \mathcal{O}(\cot\beta \lambda_6) \nonumber \\
U_{22} &= 1 + \cot^2\beta \left( -\frac{1}{2}
-\frac{\mu_8^2 v_2^2}{4(\lambda_2 v_2^2- \lambda_3 v_D^2)^2}\right)
-\lambda_6 \cot\beta \frac{\mu_8 v_D v_2^2}{2 \sqrt{2} (\lambda_2 v_2^2 - \lambda_3 v_D^2)^2}
- \lambda_6^2 \frac{v_2^2 v_D^2}{8(\lambda_2 v_2^2 - \lambda_3 v_D^2)^2} \nonumber \\
U_{23} &= - \cot\beta  \frac{\mu_8 v_2}{\sqrt{2} (\lambda_2 v_2^2- \lambda_3 v_D^2)}
- \lambda_6  \frac{ v_2 v_D}{2 (\lambda_2 v_2^2- \lambda_3 v_D^2)}
+ \mathcal{O}(\cot^2\beta) \nonumber \\
U_{33} &= 1- \cot^2\beta \frac{v_2^2}{4} \left(
\frac{ \mu_8^2}{(\lambda_2 v_2^2- \lambda_3 v_D^2)^2} + \frac{2}{v_D^2} \right)
- \lambda_6^2 \frac{v_2^2 v_D^2}{8 (\lambda_2 v_2^2 - \lambda_3 v_D^2)^2}
-  \lambda_6 \tan\beta 
\frac{\mu_8 v_D v_2^2}{2 \sqrt{2} (\lambda_2 v_2^2 - \lambda_3 v_D^2)^2} , \nonumber
\end{align}
where the $U$ matrix is approximate anti-symmetric that $U_{21} \sim - U_{12}$, $U_{31} \sim - U_{13}$ and
$U_{23} \sim - U_{32}$.

\section{CP-odd Higgs  mixing matrix and interactions}
\label{sec:CPoddHiggs}

The mass matrix of CP-odd Higgs are given in Eq.~\ref{eq:massMCPodd} and the mass of $A^0$ is given in Eq.~\ref{eq:massofCPodd}.
It is straight forward to calculate the mixing matrix $U_{\rm{odd}}$,

\begin{align}
\left(\begin{array}{c} a^0_1 \\ a^0_2 \\ a^0_3 \end{array}\right)   = U_{\rm{odd}} \left(\begin{array}{c} H_1^0 \\ H^0_2 \\ H^0_3  
\end{array}\right) = 
\left(\begin{array}{ccc}
\frac{v_D v_2}{\bar{v}^2} & \frac{v_1}{v}& - \frac{v_2}{v} \frac{v_1 v_2}{\bar{v}^2}
\\
-\frac{v_D v_1}{\bar{v}^2} & \frac{v_2}{v}& \frac{v_1}{v} \frac{v_1 v_2}{\bar{v}^2}
\\
\frac{v_1 v_2}{\bar{v}^2} & 0& \frac{v_D v}{\bar{v}^2}
\end{array}\right) 
\left(\begin{array}{c} A^0 \\ G^0_2 \\ G^0_3  
\end{array}\right),
\end{align}
where $\bar{v}^2 \equiv \left(v_1^2 v_2^2 + v_1^2 v_D^2+v_2^2 v_D^2 \right)^{1/2}$. At leading order $c_\beta$ approximation, $U_{\rm{odd}}$ becomes
\begin{align}
U_{\rm{odd}} \approx 
\left(\begin{array}{ccc}
1 - \frac{c_\beta^2 (v^2+v_D^2)}{2 v_D^2} & c_\beta & - \frac{c_\beta v}{v_D}
\\
-c_\beta & 1-c_\beta^2 &  \frac{c_\beta^2 v}{v_D}
\\
\frac{c_\beta v}{v_D} & 0 & 1 -  \frac{c_\beta^2 v^2}{2 v_D^2}
\end{array}\right) ,
\end{align}
and in this limit $G^0_2$ and $G^0_3$ are eaten by $Z$ and $K$ respectively.
The interactions between $A^0$ and bottom and top quarks are
\begin{align}
\mathcal{L}^{A^0}_{\rm tb} & = - i\frac{ m_t}{ t_\beta v} \bar{t}_L t_R A^0 
+ i \frac{m_{\tilde b}  }{  t_\beta v } 
\left(  c_{b,L}  - c_{b,R} \frac{s_\beta v^2}{v_D^2}  \right) \bar{{\tilde{b}}}_L\tilde{b}_R A^0 \\ 
& i \frac{m_{\tilde b}   }{  t_\beta v } 
\left( c_{b,L} \frac{c_{b,R}}{s_{b,R}}  +  \frac{v^2}{v_D^2} s_\beta  s_{b,R} \right) \bar{{\tilde{b}}}_L \tilde{\psi}_{b,R} A^0
\nonumber \\
&  - i \frac{m_{\tilde b}  }{  t_\beta v} 
\left( \frac{c_{b,L}}{s_{b,L}} c_{b,R} \frac{v^2}{v_D^2} s_\beta +  s_{b,L}  \right) \bar{{\tilde{\psi}}}_{b,L} \tilde{b}_{R} A^0
 \nonumber \\
& i \frac{m_{\tilde b}   }{ t_\beta v } 
\left( c_{b,L} \frac{s_{b,R}}{s_{b,L}} \frac{v^2}{v_D^2}  - c_{b,R} \frac{s_{b,L}}{s_{b,R}}  \right) 
\bar{{\tilde{\psi}}}_{b,L} \tilde{\psi}_{b,R} A^0  + h.c. \nonumber ,
\end{align}
where we see the dominant interaction is with $\bar{t}t$ only suppressed by $c_\beta$.
The interactions between $A^0$ and charm and strange quarks are similar, 
\begin{align}
\mathcal{L}^{A^0}_{\rm cs} & =  i\frac{ m_s}{t_\beta v} \bar{s}_L s_R A^0 
- i \frac{m_{\tilde c}  }{  t_\beta v } 
\left(  c_{c,L}  - c_{c,R} \frac{s_\beta v^2}{v_D^2}  \right) \bar{{\tilde{c}}}_L\tilde{c}_R A^0 \\ 
& - i \frac{m_{\tilde c}   }{  t_\beta v } 
\left( c_{c,L} \frac{c_{c,R}}{s_{c,R}}  +  \frac{v^2}{v_D^2} s_\beta  s_{c,R} \right) \bar{{\tilde{c}}}_L \tilde{\psi}_{c,R} A^0
\nonumber \\
&   i \frac{m_{\tilde c}  }{  t_\beta v} 
\left( \frac{c_{c,L}}{s_{c,L}} c_{c,R} \frac{v^2}{v_D^2} s_\beta +  s_{c,L}  \right) \bar{{\tilde{\psi}}}_{c,L} \tilde{c}_{R} A^0
 \nonumber \\
& - i \frac{m_{\tilde c}   }{ t_\beta v } 
\left( c_{c,L} \frac{s_{c,R}}{s_{c,L}} \frac{v^2}{v_D^2}  - c_{c,R} \frac{s_{c,L}}{s_{c,R}}  \right) 
\bar{{\tilde{\psi}}}_{c,L} \tilde{\psi}_{c,R} A^0  + h.c. \nonumber ,
\end{align}
where there is a minus difference between $tb$ and $cs$ quarks from $\Phi_2$ and $\tilde{\Phi}_2$. We see $A^0$ can decay to SM top, bottom, charm and strange quark pair with width proportional to $m_q^2 c_\beta^2$ in the leading terms.

\bibliography{referencelist}

\bibliographystyle{JHEP}   

\end{document}